\documentclass{elsart}

\usepackage[square,sort&compress,comma]{natbib}

\usepackage{graphicx}
\usepackage{bm}
\usepackage{amsmath}
\usepackage{amsfonts}
\newlength{\twocolumnwidth}
 
\newcommand{\preprintmargin}{ } 
\newcommand{\Eq}[1]{Eq.\ (\protect\ref{eq:#1})}
\newcommand{\Eqs}[1]{Eqs.\ (\protect\ref{eq:#1})}

\newcommand{\hide}[1]{} 
\newcommand{\eqlabel}[1]{\addtocounter{equation}{1}
	\tag{\arabic{equation}{\scriptsize \{eq:#1\}}}\label{eq:#1}} 
 
\newcommand{\chlabel}[1]{\{#1\}\label{#1}} 

\newcommand{\Heis}[1]{\protect{\hat {\mathcal #1}}} 
 
\newcommand{\R}{_{\text{R}} }

\newcommand{\av}[1]{\overline{\hspace{0.1ex}#1\hspace{0.1ex}}} 
 
\newcommand{\Tr}{\text{Tr}}

\renewcommand{\bm}[1]{{\mbox{\rm\boldmath$#1$}}}

\newcommand{\generalbracket}[4]{\protect\ensuremath{#1#2#4#1#3}} 
 
\newcommand{\Sbracket}[2]{\protect\generalbracket{#1}{[}{]}{#2}} 
 
\newcommand{\Qbracket}[2]%
 {\protect\generalbracket{#1}{\langle}{\rangle}{#2}} 
\newlength{\QQlength}
 
\newcommand{\Vbracket}[2]%
 {\protect\generalbracket{#1}{\langle 0#1|}{|0#1\rangle}{#2}} 
\newcommand{\vacavbracket}[2]%
 {#1\langle 0#1|#2#1|0#1\rangle} 
\newcommand{\A}{ }
\newcommand{\formA}[1]%
 {{\renewcommand{\A}{&}\begin{aligned}#1\end{aligned}}}
\newcommand{\formG}[1]%
 {{\renewcommand{\A}{ }\begin{gathered}#1\end{gathered}}}
\newcommand{\eqM}[2]%
 {{\renewcommand{\A}{ }\begin{multline}\preprintmargin #1\preprintmargin #2\end{multline}}}
\newcommand{\eqMW}[2]%
 {{\renewcommand{\A}{ }\begin{multline}#1 #2\end{multline}}}
\newcommand{\eqA}[2]%
 {\protect{\begin{align}{{\renewcommand{\A}{&}\begin{aligned}#1\end{aligned}}}#2\end{align}}}
\newcommand{\eqG}[2]%
 {\protect{\begin{gather}{{\renewcommand{\A}{ }\begin{gathered}#1\end{gathered}}}#2\end{gather}}}
 
\newcommand{\qav}[1]{\left\langle #1\right\rangle} 
\newcommand{\nav}[1]{\left\langle \bm{:} #1 \bm{:} \right\rangle} 
 
\newcommand{\ket}[1]{\left|#1\right\rangle} 
 
\newcommand{\vac}{\ket{0}} 
\newcommand{\melc}[3]
	{\left\langle #1\left|#2\right|#3\right\rangle}

\newcommand{\vacav}[1]{\melc{0}{#1}{0}}

\newcommand{\dg}{^{\dag}} 
\newcommand{\dgp}[1]{^{\dag #1}} 
 
\newcommand{\ULM}{Abteilung Quantenphysik, Universit\"at Ulm, 
D-89069 Ulm, Germany} 

\newcommand{\PRD}[1]{Phys.\ Rev.\ D \textbf{#1}}

\newcommand{\sP}{} 
\renewcommand{\eqlabel}[1]{\label{eq:#1}}
\renewcommand{\chlabel}[1]{\label{#1}} 
\providecommand{\flashlabel}[1]{\label{#1}} 

\newcommand{\StigAffilSw}{Physics Department, Royal Institute of Technology, KTH, Stockholm, Sweden}
\newcommand{\StigAffilFin}{Laboratory of Computational Engineering, HUT, Espoo, Finland}
\begin{document}

\begin{frontmatter}



\title{Causal signal transmission by quantum fields.\\  
I.\ Response of the harmonic oscillator}


\author[a1]{L.\ I.\ Plimak} 
\author[a1,a2,a3]{and S.\ Stenholm} 
\address[a1]{\ULM.} 
\address[a2]{\StigAffilSw.} 
\address[a3]{\StigAffilFin.} 

\begin{abstract}
It is shown that response properties of a quantum harmonic oscillator are in essence those of a classical oscillator, and that, paradoxical as it may be, these classical properties underlie all quantum dynamical properties of the system. The results are extended to non-interacting bosonic fields, both neutral and charged. 
\end{abstract}


\end{frontmatter}
\section{Introduction} 
The ultimate goal of this and forthcoming papers is to extend the phase-space techniques developed initially within the quantum-optical paradigm \cite{Schleich,MandelWolf} to the relativistic quantum field theory. 
To this end, we introduce the concept of {\em 
quantum-statistical response formulation\/} 
of an arbitrary quantum-field 
system and show its equivalence with the conventional 
Green-function techniques of the quantum field theory \cite{Bogol,Zinn,Keldysh}. 
We prove that Green's functions in the quantum field theory express, in a somewhat tangled manner, quantum-statistical response properties of the quantum system. 
In short, we show that the actual content of the quantum 
field theory is {\em microscopic signal propagation\/}. 
Response formulation employs the phase-space concepts making it a natural ``bridging'' approach between the quantum field theory and the phase-space techniques. 
We start our analyses with the simple case of a harmonic oscillator. 
Results found for this generic system then serve as leading considerations when analysing response properties of more complicated quantum systems. In this paper, we show that response formulation may be introduced for any noninteracting bosonic system, whether neutral or charged. In forthcoming papers we further extend the results to arbitrary interacting quantum fields, both bosonic and fermionic.

Response is a universal property of any physical 
system relating all other properties of the system to experimentally 
observable quantities. 
In particular, formulating a system in terms of its response  
removes the annoying incompatibility between the quantum and classical 
techniques (q-numbers {\em versus\/} c-numbers). 
A system that is {\em physically\/} classical (i.e, non-quantum in its observable properties) becomes in a response formulation also {\em formally\/} classical. 
Response reformulation of a physically classical system recovers a c-number description of the same system. This occurs the facts notwithstanding that this system has been initially formulated in q-number terms,  
and that the overall approach remains fully quantum.    
In this paper we show in detail how this ``recovery of classicality'' in quantum mechanics emerges for the simplest example of a harmonic oscillator. 
We demonstrate that response properties of a harmonic oscillator are exactly those of the classical oscillator, and that, paradoxical as it may sound, these classical response properties, in fact, 
underlie all quantum dynamical properties of the oscillator.  
A classical interpretation may be given even to the vacuum fluctuations in the harmonic oscillator which are commonly believed to be the quintessence of the quantum. 

A word of caution is in place here. The literature on the harmonic oscillator is enormous, and majority, or may be even all, of the technical points we address in this paper are included in the earlier work, see, e.g., \cite{Nelson,Vogel,QED}. What we regard as the actual result is not these formulae as such, but the pattern of response fomulation we introduce based on them. The utility of our approach will eventually emerge when we extend the treatment to interacting fields in forthcoming papers. 

Perhaps the best way of putting our results, however informally, into the right historical perspective is, \mbox{``Schwinger$=$Kubo$+$Glauber.''} Schwinger \cite{SchwingerJ} recognised the importance of formal c-number currents in the equations for Green's functions in the quantum field theory. Kubo \cite{Kubo} formulated his linear response theory by adding classical source terms to quantum Hamiltonians. Glauber \cite{GlauberPhDet} was first to understand the relevance of phase-space concepts to macroscopic measurements. Our analyses show that Schwinger's c-number currents should be kept distinct from Kubo's external sources. This is especially clear within the so-called closed-time-loop formalism by Schwinger \cite{SchwingerC}, see also \cite{Keldysh}. In this formalism, a pair of c-number Schwinger currents are introduced governing quantum evolution forwards and backwards in time. We show how one can rearrange these into two combinations, the first corresponding to the Kubo source, and the second---to the self-radiation of the system in the presence of this source. Full information on response properties of the system hidden in Schwinger's closed-time-loop formulation is then conveniently reexpressed by combining Kubo's and Glauber's approaches. Detailed discussion of these and related points will be given in section \ref{ch:Disc}.

The paper is organised as follows. 
Quantum-statistical response properties of the oscillator are analysed in section \ref{ch:DrivOsc}. Extension of these analyses to arbitrary noninteracting bosonic fields is given in section \ref{ch:LField}.  
In section \ref{ch:RespRef} we introduce a {\em response formulation\/} of noninteracting bosons and clarify its relation to Glauber's photodetection theory \cite{GlauberPhDet,KellyKleiner,GlauberTN,MandelWolf}.  
In section \ref{ch:Kubo} we discuss the relation between the response formulation and Kubo's linear response theory \cite{Kubo}. 
In section \ref{ch:Schwinger} we compare the concept of external c-number current in Kubo's and Schwinger's approaches. 
In appendix \ref{app:Wick} we outline Hori's form of Wick's theorem \cite{Hori} in the closed-time-loop formalism. 
In appendix \ref{app:Symm} we prove that an alternative way of defining the Kubo current through the  Schwinger currents discussed in \ref{ch:Schwinger} is indeed related to the symmetric operator ordering.

\section{Quantum-statistical response of the harmonic oscillator}%
\chlabel{ch:DrivOsc} 
\subsection{Harmonic oscillator basics} 
We consider the quantum harmonic oscillator, 
with the Hamiltonian in the Schr\"odinger picture 
\begin{gather} 
\eqlabel{OscHam0} 
\Heis{H}_{0\text{S}}   = \frac{\hat p^2}{2m} + \frac{m\omega_0 ^2\hat q ^2}{2}
= \hbar \omega_0 \left(
\hat a\dg \hat a +\frac{1}{2} 
\right). 
\end{gather}%
Here $\hat q$ and $\hat p$ are the standard position and momentum 
operators 
\begin{align} 
\begin{aligned}
\left[
\hat q, \hat p
\right] = i \hbar   , 
\end{aligned}%
\end{align}%
and $\hat a\dg,\hat a$ are the usual creation and annihilation operators, 
\begin{gather} 
\hat a = \frac{1}{\sqrt{2}} \left(
\frac{\hat q}{q_0} - i\, \frac{\hat p}{p_0} 
\right) , \ \ 
\left[
\hat a, \hat a\dg 
\right] =1
 , 
\eqlabel{ADef} 
\\  
q_0 = \sqrt\frac{\hbar }{m\omega_0 } 
 , \ \  
p_0 = \sqrt{\hbar m\omega_0 } . 
\eqlabel{p0q0} 
\end{gather}%
Furthermore, the vacuum (ground) state of the oscillator is defined by 
\begin{align} 
\begin{aligned}
\hat a \vac = 0 . 
\end{aligned}%
\end{align}%
Assuming that 
the Schr\"odinger and Heisenberg pictures coinside at $t_0=0$, 
it is easy to find the Heisenberg operators related to the Hamiltonian (\ref{eq:OscHam0}):  
\begin{gather} 
\begin{gathered} 
\hat a(t) = \hat a \textrm{e}^{-i\omega_0 t}, 
\ \ \hat a\dg(t) = \hat a\dg \textrm{e}^{i\omega_0 t}, \\ 
\hat q(t) = \frac{q_0}{\sqrt{2}} \big[\hat a(t)  + \hat a\dg(t) \big], \\
\hat p(t) = \frac{i p_0}{\sqrt{2}} \big[\hat a(t)  - \hat a\dg(t) \big], 
\end{gathered}%
\eqlabel{OscHeis} 
\end{gather}%
where 
$\hat a(t) = \textrm{e}^{-(i/\hbar ) \hat H_0 t}\hat a\textrm{e}^{(i/\hbar ) \hat H_0 t}$,  
and similarly for $\hat q(t)$ and $\hat p(t)$. 
Finally, the {\em normal ordering\/} 
places all $\hat a\dg$s to 
the left of all $\hat a$s, e.g. 
\eqA{ 
\bm{:} \hat q(t) \hat q(t') \bm{:} = 
\frac{q_0^2}{2} \big[
\hat a(t) \hat a(t') + 
\hat a\dg(t) \hat a(t') + 
\hat a\dg(t') \hat a(t) + 
\hat a\dg(t) \hat a\dg(t') 
\big] 
.  
}{
}%
\subsection{Driven oscillator} 
Following Kubo \cite{Kubo}, we analyse response of the oscillator by adding a source term to the Hamiltonian (\ref{eq:OscHam0}) 
\begin{gather} 
\begin{aligned}
\Heis{H}_{\text{S}}(t) = \Heis{H}_{0\text{S}} + j(t)\hat q . 
\end{aligned}%
\eqlabel{HamJ} 
\end{gather}%
For simplicity we assume that $j(t)=0,\ t<0$. 
We consider the source term as a perturbation, but do not assume it to be small. In this terminology, $\Heis{H}_{0\text{S}}$ is a free Hamiltonian and $\Heis{H}_{\text{S}}(t)$ is a full Hamiltonian, while $\hat q(t)$ and $\hat p(t)$ are the interaction-picture operators. 
The Heisenberg position and momentum operators related to $\Heis{H}_{\text{S}}(t)$ will be denoted as $\hat q_j(t)$ and $\hat p_j(t)$. 
The presense of the source term in (\ref{eq:HamJ}) makes transition to the Heisenberg picture nontrivial. The Schr\"odinger and the Heisenberg representations of an arbitrary operator 
($\hat x_{\text{S}} (t)$ and $\hat x_j (t)$, respectively) are coupled by the unitary evolution operator, 
\eqA{ 
\hat x_j (t) = \Heis{U}\dg(t)\hat x_{\text{S}} (t)\Heis{U}(t).  
}{
}%
The evolution operator obeys the Schr\"odinger equation, 
\eqA{ 
i \hbar{\dot{{\Heis{U}}}}(t) = \Heis{H}_{0\text{S}}(t) \Heis{U}(t), 
}{
}%
while $\hat x_j (t)$ obeys the Heisenberg equation 
\eqA{ 
i \hbar\dot{\hat x}_j (t) = \left[
{\hat x}_j(t), \Heis{H}_j(t)
\right] + \Heis{U}\dg(t)\dot{\hat x}_{\text{S}} (t)\Heis{U}(t),
}{
}%
where $\Heis{H}_j(t)$ is the full Hamiltonian 
in the Heisenberg picture,
\eqA{ 
\Heis{H}_j(t) = \Heis{U}\dg(t)\Heis{H}_{0\text{S}}(t)\Heis{U}(t) = 
\frac{\hat p^2_j(t)}{2m} + \frac{m\omega_0 ^2\hat q^2_j(t) }{2} 
+ j(t) \hat q_j(t). 
}{
}%
In particular, 
\begin{gather} 
\begin{aligned}
i\hbar \dot {\hat q}_j(t) = \left[
{\hat q}_j(t), \Heis{H}_j(t)
\right] ,  \\ 
i\hbar \dot {\hat p}_j(t) = \left[
{\hat p}_j(t), \Heis{H}_j(t)
\right] .   
\end{aligned}%
\eqlabel{HeisQP} 
\end{gather}%
Commutators in (\ref{eq:HeisQP}) are easily calculated resulting in the classically-looking pair 
\begin{gather} 
\begin{aligned}
\dot {\hat q}_j(t) = \frac{{\hat p}_j(t)}{m}, \ \ 
\dot {\hat p}_j(t) = - m\omega ^2_0 {\hat q}_j(t) - j(t) 
. 
\end{aligned}%
\end{gather}%
Substituting $\dot {\hat p}_j(t) = m \ddot {\hat q}_j(t)$ into the second equation results in the inhomogeneous ``wave'' equation for the oscillator displacement 
\begin{gather} 
\begin{aligned}
\ddot {\hat q}_j(t) + \omega ^2_0 {\hat q}_j(t) = - \frac{j(t)}{m} ,  
\end{aligned}%
\end{gather}%
while ${\hat q}(t)$ obeys the homogeneous equation 
\begin{gather} 
\begin{aligned}
\ddot {\hat q}(t) + \omega ^2_0 {\hat q}(t) = 0 .   
\end{aligned}%
\end{gather}%
By noting that ${\hat q}_j(t) = {\hat q}(t)$ for $t<0$ we find 
\begin{gather} 
\begin{aligned}
\hat q_j(t) = \hat q(t)+ q_j(t) ,  
\end{aligned}%
\eqlabel{qjbyq} 
\end{gather}%
where 
\begin{align} 
\begin{aligned}
q_j(t) = \int dt' D_{\text{R}}(t-t') j(t') . 
\end{aligned}%
\eqlabel{qj} 
\end{align}%
Omitting limits of a time integration means that this integration is from plus to minus infinity, cf.\ endnote \cite{IntRange}.  
In (\ref{eq:qj}), $D_{\text{R}}$ is, up to overall sign, the retarded Green's function of the  {\em c-number\/} equation for the displacement of a classical oscillator under the influence 
of an external force  $f(t) = -j(t)$ 
\begin{align} 
\begin{gathered} 
\ddot {{q}}(t) + \omega_0 ^2 {{q}}(t) = \frac{f(t)}{m} , \\ 
q(t) = q_{\text{in}} (t)- \int dt' D_{\text{R}} (t-t')f(t')
, 
\end{gathered}%
\eqlabel{OscEqQCl} 
\end{align}%
where $q_{\text{in}} (t)$ is a solution with $j=0$.
The kernel $D_{\text{R}}$ obeys the typical inhomogeneous ``wave'' equation for Green's function, 
\begin{align} 
\begin{aligned}
& \ddot D_{\text{R}} (\tau ) + \omega_0 ^2 D_{\text{R}} (\tau ) = 
- \frac{\delta(\tau )}{m}  
.  
\end{aligned}%
\eqlabel{OscEqDret} 
\end{align}%
Explicitly, for the harmonic oscillator (\ref{eq:OscHam0}) we have
\begin{align} 
\begin{aligned}
D_{\text{R}} (t) = -\theta (\tau ) \frac{\sin \omega_0 \tau }{m\omega_0 }. 
\end{aligned}%
\eqlabel{OscDRExpl} 
\end{align}%

Equation (\ref{eq:qjbyq}) has an important consequence. The difference between $\hat q_j(t)$ and $\hat q(t)$ is a c-number and does not interfere with operator orderings. The concept of normal ordering defined for the interaction-picture operators $\hat q(t)$ and $\hat p(t)$ thus equally applies to the Heisenberg-picture operators $\hat q_j(t)$ and $\hat p_j(t)$. 
\subsection{Quantum-field-theoretical approach to the driven oscillator} 
\newcommand{\Q}[1]{\hat q(t #1)} 
\newcommand{\Qj}[1]{\hat q_j(t #1)} 
\newcommand{\etp}[1]{\eta_+(t #1)} 
\newcommand{\etm}[1]{\eta_-(t #1)} 
The goal of our analyses is to find a way of approaching interacting nonlinear quantum systems. 
To this end, we treat the oscillator using concepts which are normally reserved for much heavier problems, where a solution in terms of operators is a rare exception. 
Namely, we employ the well-known {\em closed-time-loop\/} Schwinger-Perel-Keldysh formalism \cite{SchwingerC,Keldysh}, applicable to any nonlinear nonequilibrium quantum-statistical problem (see, e.g., a review by Fred Cooper \cite{Cooper}). 
In this formalism, the oscillator is characterised in terms of the {\em double-time-ordered\/} averages of the displacement operators 
\eqM{ 
\qav{T_-\Q{_1}\cdots\Q{_m}\,
T_+\Q{'_1}\cdots\Q{'_n}} \\ = 
\frac{(-i)^m i^n \, 
\delta^{m+n}\Phi (\eta _-,\eta _+)}{
\delta\etm{_1} 
\cdots 
\delta\etm{_m} 
\delta\etp{'_1} 
\cdots 
\delta\etp{'_n} 
} 
\Big|_{\eta _-=\eta _+ =0}
\, 
,  
}{
\eqlabel{FGNL} 
}%
where we have introduced the generating, or characteristic,  
functional 
\begin{multline} 
\Phi (\eta _-,\eta _+) = 
\bigg\langle
T_- \exp \bigg [ i \int dt\,\eta _-(t) {\hat{{q}}}(t)\bigg ] 
\ T_+ \exp \bigg [-i \int dt\,\eta _+(t) {\hat{{q}}}(t)\bigg ]
\bigg\rangle \\ = 1+  
\sum_{m+n\geq 1} \frac{i^m (-i)^n }{m!n!} 
\left(\int dt\right)^{m+n} 
\eta _-(t_1)\cdots\eta _-(t_m)
\eta _+(t'_1)\cdots\eta _+(t'_n) \\ \times 
\qav{T_-{\hat q}(t_1)\cdots{\hat q}(t_m)
T_+{\hat q}(t'_1)\cdots{\hat q}(t'_n)}
,     
\eqlabel{PhiDef} 
\end{multline}%
with the functional arguments  
$\eta _{\pm}(t)$ 
being arbitrary smooth c-number functions. 
The symbol $\left(
\int dt
\right)^{m+n} $ denotes integration over all time arguments, cf.\ also endnote \cite{IntRange}. 
Time-ordering $T_+$ ($T_-$) puts operators in order of 
increasing (decreasing) time arguments. That is, if $t_1<t_2<\cdots<t_m$, 
\begin{gather} 
\begin{aligned}
&T_-{\hat q}(t_1)\cdots{\hat q}(t_m)  =  
{\hat q}(t_1)\cdots{\hat q}(t_{m-1}){\hat q}(t_m), 
\\
&T_+{\hat q}(t_1)\cdots{\hat q}(t_m)  = 
{\hat q}(t_m){\hat q}(t_{m-1})\cdots{\hat q}(t_1) 
.   
\end{aligned}%
\eqlabel{TpmDef} 
\end{gather}%
By definition, bosonic operators under the ordering sign commute. 
The quantum averaging in 
(\ref{eq:FGNL}), (\ref{eq:PhiDef}) is defined in the standard way 
with respect to the initial state of the oscillator, described by the density matrix $\rho _0$, 
\begin{gather} 
\begin{aligned}
\qav{\cdots} = \Tr \rho_0 (\cdots). 
\end{aligned}%
\eqlabel{QAv0} 
\end{gather}%
Functional (\ref{eq:PhiDef}) obeys the reality property, 
\begin{gather} 
\begin{aligned}
\Phi^*  (\eta _-,\eta _+) = \Phi  (\eta _+^*,\eta _-^*), 
\end{aligned}%
\eqlabel{PhiC} 
\end{gather}%
which is a consequence of 
\begin{gather} 
\begin{aligned}
{[T_{\pm}\hat x(t)\cdots\hat y(t')]}\dg = T_{\mp}\hat x\dg(t)\cdots\hat y\dg(t'), 
\end{aligned}%
\end{gather}%
where $\hat x(t)\cdots\hat y(t')$ are arbitrary operators. 
This formula expresses the simple fact that Hermitian conjugation inverts the order of factors turning a $T_+$-ordered product into a $T_-$-ordered one, and {\em vice versa\/}.

\begin{figure}[t]
\begin{center}
\includegraphics[width=0.65\columnwidth]{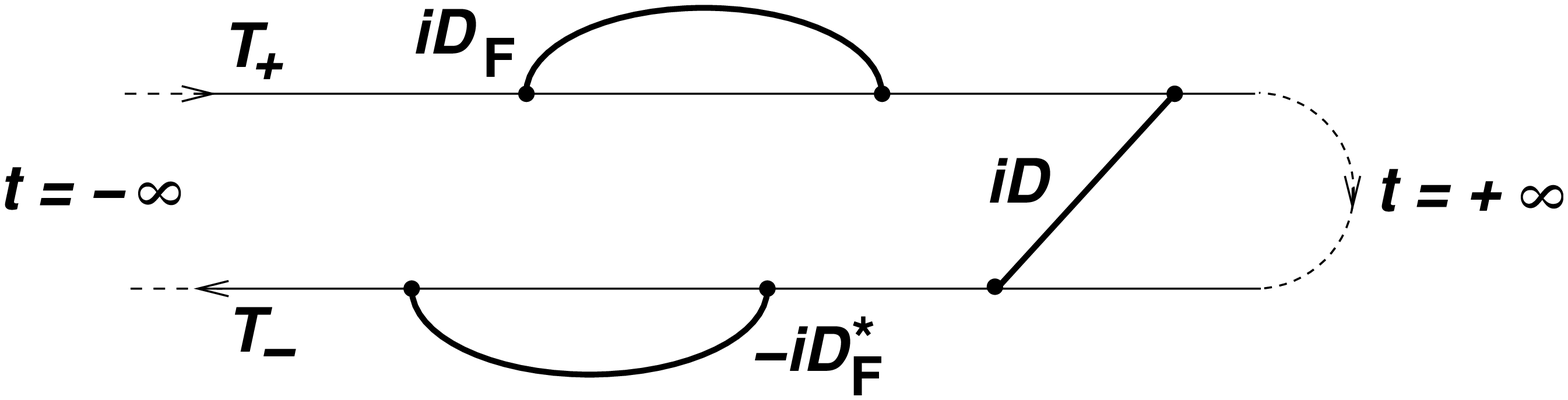}
\end{center}
\caption{%
C-contour (thin lines) travels first from $t=-\infty$ to $t=+\infty$ (direct branch) and then back to $t=-\infty$ (reverse branch). All operators on the direct branch are placed to the right of all operators on the reverse branch, while operators on the direct (reverse) branch are $T_+$ ($T_-$) ordered among themselves. This ordering rule obviously reproduces the double-time ordering defined by \Eq{FGNL}. The three contractions defined by \Eqs{DFDef}--(\ref{eq:DDef}) are shown schematically as thick lines. They correspond to three possible ways for an operator pair to be positioned on the C-contour: both operators on the direct branch, both operators on the reverse branch, and one operator on the direct branch, the other on the reverse branch.}
\label{fig:C}
\end{figure}
The double time ordering defined by \Eq{FGNL} is often understood as operator ordering on the so-called {\em C-contour\/} shown in Fig.\ \ref{fig:C}. It travels first from $t=-\infty$ to $t=+\infty$ and then back to $t=-\infty$ (hence the name ``closed-time-loop formalism''). All operators on the direct branch are placed to the right of all operators on the reverse branch, while operators on the direct (reverse) branch are $T_+$ ($T_-$) ordered among themselves. This ordering rule obviously reproduces the double time ordering defined by \Eq{FGNL} --- note that ``earlier'' operators go to the right.  
However, for our purposes it is more convenient to keep the $T_{\pm}$-orderings explicit. 
The interpretation of $\Phi (\eta _-,\eta _+)$ as Schwinger's closed-time-loop amplitude \cite{SchwingerC} will be discussed in \ref{ch:Schwinger}.

Definitions identical to (\ref{eq:FGNL}), (\ref{eq:PhiDef}) are also given for the quantities related to operator $\hat q_j(t)$:  
\begin{multline} %
\preprintmargin
\qav{T_-\Qj{_1}\cdots\Qj{_m}\,
T_+\Qj{'_1}\cdots\Qj{'_n}} \\ = 
\frac{(-i)^m i^n \, 
\delta^{m+n}\Phi (\eta _-,\eta _+;j)}{
\delta\etm{_1} 
\cdots 
\delta\etm{_m} 
\delta\etp{'_1} 
\cdots 
\delta\etp{'_n} 
} 
\Big|_{\eta _-=\eta _+ =0}
\, 
,    
\eqlabel{PhiDefJ} 
\preprintmargin
\end{multline}%
where 
\begin{multline} %
\preprintmargin
\Phi (\eta _-,\eta _+;j) \\ = 
\bigg\langle
T_- \exp \bigg [ i \int dt\,\eta _-(t) {\hat{q}}_j(t)\bigg ] 
T_+ \exp \bigg [-i \int dt\,\eta _+(t) {\hat{q}}_j(t)\bigg ]
\bigg\rangle .  
\eqlabel{FGNLJ} %
\preprintmargin
\end{multline}%
Equation (\ref{eq:qjbyq}) then results in a simple formula expressing $\Phi (\eta _-,\eta _+;j)$ by $\Phi (\eta _-,\eta _+)$: 
\begin{gather} 
\begin{aligned}
\Phi (\eta _-,\eta _+;j) = \Phi (\eta _-,\eta _+)\Phi_{\text{cl}}  (\eta ;j ) . 
\end{aligned}%
\eqlabel{PhiJ} 
\end{gather}%
Here, 
\begin{gather} 
\begin{aligned}
\eta (t) = -i \left[
\eta _+(t)- \eta_- (t)
\right] ,  
\end{aligned}%
\eqlabel{EtaDef} 
\end{gather}%
and 
\begin{gather} 
\begin{gathered} 
\Phi_{\text{cl}}  (\eta ;j ) = \exp \int dt\, \eta (t) q_j(t), 
\end{gathered}%
\eqlabel{Fcl} 
\end{gather}%
where $q_j(t)$ given by \Eq{qj} is the classical displacement caused by the source $j(t)$. The quantity $\Phi_{\text{cl}}  (\eta ;j )$ is nothing but the characteristic functional for products of these displacements at different times: 
\begin{gather} 
\begin{aligned}
q_j(t_1)\cdots q_j(t_m) =  
\frac{\delta^m \Phi_{\text{cl}}  (\eta ;j )}
{\delta\eta (t_1)\cdots \delta\eta (t_m)}\Big|_{\eta =0} . 
\end{aligned}%
\end{gather}%
Obviously, $\Phi_{\text{cl}}  (\eta ;j )$ is a purely classical object.

Equation (\ref{eq:PhiJ}) is a good example of how manipulating charactestic functionals gives access to physics contained in quantum Green's functions. It is also the first indication that merely by introducing new functional arguments one can make this physics much more transparent. 
\subsection{Wick's theorem and 
vacuum Green's function} 
Relation (\ref{eq:PhiJ}) expresses two well-known properties of the harmonic oscillator: that the classical external source creates a coherent amplitude, 
and that intrinsic quantum properties and response of a linear system 
are independent, hence the factorisation. 
To calculate $\Phi (\eta _-,\eta _+)$, we make use of another case of quantum-statistical independence and ensuing factorisation characteristic of linear systems. Namely, 
\begin{gather} 
\begin{aligned}
\Phi (\eta _-,\eta _+) = \Phi_{\text{vac}}  (\eta _-,\eta _+) 
\Phi_{\text{in}}  (\eta) , 
\end{aligned}%
\eqlabel{InVac} 
\end{gather}%
where  
\begin{gather} 
\begin{aligned}
\Phi_{\text{in}}  (\eta)  = \qav{\bm{:} 
\exp \int dt\, \eta (t)
{\hat{{q}}}(t)
\bm{:}}  
\end{aligned}%
\end{gather}%
characterises the {\em initial state\/} of the oscillator, and  
\begin{multline} %
\preprintmargin
\Phi _{\text{vac}} (\eta _-,\eta _+) \\ = 
\bigg\langle 0\bigg|
T_- \exp \bigg [ i \int dt\,\eta _-(t) {\hat{{q}}}(t)\bigg ] 
\, 
\, T_+ \exp \bigg [-i \int dt\,\eta _+(t) {\hat{{q}}}(t)\bigg ]
\bigg| 0 \bigg\rangle ,  
\eqlabel{OscCharFVac} 
\preprintmargin
\end{multline}%
is the functional 
$\Phi  (\eta _-,\eta _+)$ for the oscillator in a {\em vacuum state\/}. 
The functional variable $\eta(t) $ in (\ref{eq:InVac}) is the same as appearing in \Eq{PhiJ} and defined by (\ref{eq:EtaDef}).

To prove equation (\ref{eq:InVac}) it suffices to note that it is nothing but a closed form of Wick's theorem for the double-time ordering. 
To clarify this connection we look for the explicit expression for $\Phi _{\text{vac}} (\eta _-,\eta _+)$. 
Referring the reader for details to appendix \ref{app:Wick}, here we proceed by noting that 
the harmonic 
oscillator in a vacuum state is a {\em Gaussian\/} system, and, 
as for any Gaussian system, 
the functional $\Phi_{\text{vac}}$ must be an exponent 
of a quadratic form, 
\eqMW{ 
\Phi_{\text{vac}}  (\eta _-,\eta _+) = 
\exp\bigg\{
i\hbar \int dt dt' 
\bigg[
- \frac{1}{2}\, \eta _+(t)\eta _+(t')D_F(t-t') 
\\ + \frac{1}{2}\, \eta _-(t)\eta _-(t')D_F^*(t-t') 
+ \eta _-(t)\eta _+(t')D(t-t')
\bigg]
\bigg\} ,
}{
\eqlabel{OscPhiVac} 
}%
where notice was also taken of \Eq{PhiC}. 
It is easy to see that the kernels here are the ``contractions,'' 
i.e., the differences between the time and 
normally ordered products of pairs of operators. Indeed, by using \Eq{OscPhiVac} to calculate pair averages we have:  
\begin{multline} %
\preprintmargin
i\hbar D_F(t-t') = 
\left.
- \frac{\delta^2\Phi_{\text{vac}}  (\eta _-,\eta _+)}
{\delta\eta _+(t)\delta\eta _+(t')} 
\right |_{\eta _+=\eta _-=0} 
\\ = 
\vacav{T_+ {\hat{q}}(t){\hat{q}}(t')}
= 
T_+ \hat q(t) \hat q(t') - 
\bm{:} \hat q(t) \hat q(t') \bm{:} 
, 
\eqlabel{DFDef} 
\preprintmargin
\end{multline}%
\begin{multline} %
\preprintmargin
- i\hbar D^*_F(t-t') = 
\left.
- \frac{\delta^2\Phi_{\text{vac}}  (\eta _-,\eta _+)}
{\delta\eta _-(t)\delta\eta _-(t')} 
\right |_{\eta _+=\eta _-=0} 
\\ = 
\vacav{T_- {\hat{q}}(t){\hat{q}}(t')}
= 
T_- \hat q(t) \hat q(t') - 
\bm{:} \hat q(t) \hat q(t') \bm{:} 
, 
\eqlabel{DFCDef} 
\preprintmargin
\end{multline}%
\begin{multline} %
\preprintmargin
i\hbar D(t-t') = 
\left.
\frac{\delta^2\Phi_{\text{vac}}  (\eta _-,\eta _+)}
{\delta\eta _-(t)\delta\eta _+(t')} 
\right |_{\eta _+=\eta _-=0} 
\\ = 
\vacav{{\hat{q}}(t){\hat{q}}(t')}
= 
\hat q(t) \hat q(t') - 
\bm{:} \hat q(t) \hat q(t') \bm{:} 
. 
\eqlabel{DDef} 
\preprintmargin
\end{multline}%
The contractions, which in the quantum field theory are often 
called free-field Green's functions, are expressed 
by commutators and are therefore c-numbers, while 
their expressions as vacuum averages  
follow from $\vacav{\bm{:} \hat q(t) \hat q(t') \bm{:}}=0$  
(which holds  
for any normally ordered product). 
In the Feynman diagram techniques \cite{Bogol,Zinn} $D_F$ emerges as a 
propagator, while $D_F^*$ and $D$ 
occur also as parts of the matrix propagator 
in the Perel-Keldysh techniques \cite{Keldysh}. 
By making use of \Eqs{OscHeis}, for the harmonic oscillator we find
\eqA{ 
D(t) \A = - \frac{i\,\textrm{e}^{-i \omega_0 t}}{2 m \omega_0 } ,  \A 
D_F(t) \A= \theta(t) D(t)  + \theta(-t) D(-t)   
. 
}{
\eqlabel{OscDDFExpl} 
}%

We see that 
equation (\ref{eq:InVac}) has ``Wick's theorem'' written all over it. Its LHS is a characteristic functional of the double-time-ordered operator averages. Its RHS contains the contractions and normally-ordered operator averages. The three contractions defined by \Eqs{DFDef}--(\ref{eq:DDef}) correspond to the three ways for an operator pair to be placed on the C-contour: both operators on the direct branch, both operators on the reverse branch, and one operator on the direct branch, the other on the reverse one, cf.\ Fig.\ \ref{fig:C}. On expanding both sides of (\ref{eq:InVac}) into functional Taylor series and comparing the coefficients, we find expressions for the double time-ordered operator averages in terms of sums of the normal averages with contractions, just as stipulated by Wick's theorem. The remaining tedious task of verifying that all terms in these sums indeed occur with coefficient 1 may be replaced by a direct algebraic proof of (\ref{eq:InVac}) based on Hori's form of Wick's theorem \cite{Hori}. For details of this proof we refer the reader to appendix \ref{app:Wick}.  
\subsection{Are vacuum fluctuations quantum?} 
\subsubsection{Algebra of Green's functions}\chlabel{ch:OscAlg} 
By combining (\ref{eq:PhiJ}) with (\ref{eq:InVac}) we find 
\begin{gather} 
\begin{aligned}
\Phi (\eta _-,\eta _+ ; j) = 
\Phi_{\text{vac}}  (\eta _-,\eta _+) 
\Phi_{\text{in}}  (\eta) 
\Phi _{\text{cl}}  (\eta ;j) 
.   
\end{aligned}%
\eqlabel{FFF} 
\end{gather}%
Quantum properties of the driven oscillator thus comprise three independent contributions: the vacuum fluctuations (characterised by $\Phi_{\text{vac}}$), the initial state (characterised by $\Phi_{\text{in}}$), and the external source (characterised by $\Phi_{\text{cl}} $). Of the three contributions, that of the external source is fully classical. Contribution of the intial state is quantum or classical depending on whether this state is quantum or classical. 
As to the vacuum fluctuations, these are commonly regarded as the quintessence of the quantum. 
However, contrary to the common wisdom, we are going to show that $\Phi_{\text{vac}}$ allows for a fully classical interpretation in terms of the response of the classical oscillator. 

As a first step to demonstrate this, we express $D_F(t)$ and $D(t)$ by $D_{\text{R}} (t)$. 
It is easy to verify that the kernel $D_{\text{R}}(t)$ 
in the quantum treatment may be defined  
as the retarded combination 
\begin{align} 
\begin{aligned}
D_{\text{R}} (t) = D_F(t) -  D(-t) = \theta(t)\big[D(t) - D(-t)\big].  
\end{aligned}%
\eqlabel{OscDretbyDDF} 
\end{align}%
From this equation we find  
\begin{multline} %
\preprintmargin
D\R(t)-D\R(-t) = 
\theta(t)\big[D(t) - D(-t)\big] - 
\theta(-t)\big[D(-t) - D(t)\big] \\ = D(t) - D(-t)
. 
\eqlabel{DDTDRDRT} 
\preprintmargin
\end{multline}%
Physics behind \Eqs{OscDretbyDDF} and (\ref{eq:DDTDRDRT}) is discussed in  
section \ref{ch:Kubo}. 
Here we proceed formally by noting that $D(t)$ may be expressed by $D\R(t)$ if retaining only {\em half the frequency spectrum\/} of \Eq{DDTDRDRT}.
Indeed, in the terminology of quantum optics, $D(t)$ is purely {\em frequency-positive\/}, $D(t)\propto \textrm{e}^{-i \omega_0 t}$, so that $D(-t)$ is purely {\em frequency-negative\/}, $D(-t)\propto \textrm{e}^{i \omega_0 t}$.  
In general, the frequency-positive 
and negative parts 
$g^{(\pm)}(t)$ of 
some function $g(t)$ are defined by dropping the corresponding half 
(negative or positive) of its Fourier-spectrum 
\begin{align} 
\begin{aligned}
g^{(\pm)}(t) = \int \frac{d\omega }{2 \pi } \,
\textrm{e}^{-i\omega t}\theta(\pm \omega )g_{\omega } , 
\ \   
g_{\omega } = \int dt\, \textrm{e}^{i\omega t} g(t).   
\end{aligned}%
\eqlabel{FPNDef} 
\end{align}%
Obviously, 
\begin{align} 
\begin{aligned}
g^{(+)}(t)+g^{(-)}(t) = g(t). 
\end{aligned}%
\eqlabel{FPNSum} 
\end{align}%
This relation, however, implies that $g_{\omega }$ is smooth at \mbox{$\omega = 0$}. 

Retaining only the frequency-positive part of (\ref{eq:DDTDRDRT}) 
yields
\begin{align} 
\begin{aligned}
D(t) = 
D^{(+)}_{\text{R}}(t) - D^{(-)}_{\text{R}}(-t) , 
\end{aligned}%
\eqlabel{OscDbyDR} 
\end{align}%
where $D^{(+)}_{\text{R}}(t)$ and $D^{(-)}_{\text{R}}(t)$ denote the 
frequency-positive and negative parts of $D_{\text{R}}(t)$, respectively.  
In obtaining \Eq{OscDbyDR} we have used the fact that inverting the time argument 
swaps the frequency-positive and 
negative parts of a function. 
Finding $D_{\text{F}} $ is now a matter of straightforward algebra. 
Again using (\ref{eq:OscDretbyDDF}), this time to express $D_{\text{F}} $, 
we have 
\begin{multline} %
\preprintmargin
D_{\text{F}} (t) = D\R(t) + D(-t) = 
D\R(t) + D^{(+)}_{\text{R}}(- t) - D^{(-)}_{\text{R}}(t) \\ = 
D^{(+)}\R(t) + D^{(+)}_{\text{R}}(- t) 
, 
\eqlabel{OscDFbyDR} 
\preprintmargin
\end{multline}%
where use was also made of \Eq{FPNSum}. 
\subsubsection{Response substitution} 
In the common parlance, \Eq{OscPhiVac} 
embodies {\em vacuum fluctuations\/} 
in the oscillator. 
However, in view of relations (\ref{eq:OscDbyDR}) 
and (\ref{eq:OscDFbyDR}), how quantum are these? 
We note that $D_{\text{R}} $ is real; for a real function, complex conjugation swaps its frequency-positive and negative parts, 
\begin{gather} 
\begin{aligned}
D\R^{(\pm)*}(t) = D\R^{(\mp)}(t). 
\end{aligned}%
\end{gather}%
Complex-conjugating (\ref{eq:OscDFbyDR}) then allows us to express $D^*_F(t)$ occuring in (\ref{eq:OscPhiVac}) as 
\begin{align} 
\begin{aligned}
D_F^*(t) = D^{(-)}_{\text{R}}(t) + D^{(-)}_{\text{R}}(-t). 
\end{aligned}%
\eqlabel{OscDFStar} 
\end{align}%
Consider now the first term in \Eq{OscPhiVac}. 
By making use of \Eq{OscDFbyDR} we have 
\begin{multline} %
\preprintmargin
i\hbar \int dt dt' 
\Big[
- \frac{1}{2} \eta _+(t)\eta _+(t')D_F(t-t') 
\Big] \\ =
- \frac{i\hbar }{2}\int dt dt' \eta _+(t)\eta _+(t') 
\Big[
D^{(+)}_{\text{R}}(t-t') + D^{(+)}_{\text{R}}(t'-t) 
\Big] \\ = 
- i\hbar \int dt dt' \eta _+(t)D^{(+)}_{\text{R}}(t-t')\eta _+(t') 
. 
\eqlabel{OscPropPP0} 
\preprintmargin 
\end{multline}%
We now employ the following connection between 
the frequency-positive and negative 
parts and the convolution: 
\begin{gather} 
\int dt' h^{(\pm)}(t-t') g(t') = 
\int dt' h(t-t') g^{(\pm)}(t') .  
\eqlabel{FPNConv} 
\end{gather}%
This is proved by noting that the Fourier-transform of either expression 
is $\theta(\pm \omega )g_{\omega }h_{\omega }$. 
From \Eqs{OscPropPP0} and (\ref{eq:FPNConv}) we find 
\eqM{ 
i\hbar \int dt dt' 
\Big[
- \frac{1}{2} \eta _+(t)\eta _+(t')D_F(t-t') 
\Big] 
\\ = - i\hbar \int dt dt' \eta _+(t)D_{\text{R}}(t-t')\eta^{(+)} _+(t') 
.  
}{
\eqlabel{OscPropPP} 
}%
Similarly, by making use of \Eqs{OscDbyDR},   
(\ref{eq:OscDFStar}) and (\ref{eq:FPNConv}),  
we transform the two remaining terms in \Eq{OscPhiVac} as follows:  
\begin{multline} %
i\hbar \int dt dt' 
\Big[
+ \frac{1 }{2} \eta _-(t)\eta _-(t')D^*_F(t-t') 
\Big] \\ =
\frac{i\hbar }{2}\int dt dt' \eta _-(t)\eta _-(t') 
\Big[
D^{(-)}_{\text{R}}(t-t') + D^{(-)}_{\text{R}}(t'-t) 
\Big] \\ = 
i\hbar \int dt dt' \eta _-(t)D^{(-)}_{\text{R}}(t-t')\eta _-(t') 
= i\hbar \int dt dt' \eta _-(t)D_{\text{R}}(t-t')\eta^{(-)} _-(t') 
, 
\eqlabel{OscPropMM} 
\end{multline}%
\begin{multline} %
i\hbar \int dt dt' 
\Big[
+ \eta _-(t)\eta _+(t')D(t-t') 
\Big] \\ = 
{i\hbar }\int dt dt' \eta _-(t)\eta _+(t') 
\Big[
D^{(+)}_{\text{R}}(t-t') -
D^{(-)}_{\text{R}}(t'-t)  
\Big] 
\\ = 
i\hbar \int dt dt' \eta _-(t)D^{(+)}_{\text{R}}(t-t')\eta _+(t') 
- 
i\hbar \int dt dt' \eta _+(t)D^{(-)}_{\text{R}}(t-t')\eta _-(t') 
\\ =
i\hbar \int dt dt' \eta _-(t)D_{\text{R}}(t-t')\eta^{(+)} _+(t') 
- i\hbar \int dt dt' \eta _+(t)D_{\text{R}}(t-t')\eta ^{(-)}_-(t')  
. 
\eqlabel{OscPropPM} 
\end{multline}%
Combining \Eqs{OscPropPP}, (\ref{eq:OscPropMM}) and (\ref{eq:OscPropPM}) 
yields 
\eqM{ 
\Phi_{\text{vac}}(\eta _-,\eta _+) 
\\ = 
\exp\left\{
-i\hbar\int dt dt' \Big[ \eta _+(t) - \eta _-(t)\Big] 
D_{\text{R}}(t-t')   \Big[
\eta^{(+)} _+(t') + \eta^{(-)} _-(t')  
\Big]
\right\} . 
}{
\eqlabel{PhiVac0} 
}%
On introducing a new set of functional variables, 
\begin{align} 
\begin{aligned}
\eta (t) & = -i\big[\eta _+(t)-\eta _-(t)\big], \\  
\sigma(t) & = \hbar \left[
\eta^{(+)} _+(t) + \eta^{(-)} _-(t)
\right]  , 
\end{aligned}%
\eqlabel{OscSubstEta} 
\end{align}%
we can write (\ref{eq:PhiVac0}) as  
\begin{gather} 
\Phi_{\text{vac}}(\eta _-,\eta _+) 
=  \exp\left[
\int dt\, \eta (t) \int dt'  
D_{\text{R}}(t-t')\sigma(t')
\right]
 .  
\eqlabel{OscPhivacResp0} 
\end{gather}%
Note that the functional variable $\eta (t)$ of \Eqs{PhiJ} and (\ref{eq:InVac}) has occurred once again. 

Equation (\ref{eq:OscPhivacResp0}), we remind the reader, 
expresses vacuum fluctuations of the 
quantum harmonic oscillator, and yet it looks astonishingly classical. 
In the variables $\eta ,\sigma $ the functional $\Phi_{\text{vac}}$ 
has turned simply into a generating functional for products of a c-number field 
emitted by c-number source $\sigma $: 
\begin{gather} 
\Phi_{\text{vac}}(\eta _-,\eta _+) 
= \Phi_{\text{cl}}  (\eta ;\sigma  )
 ,   
\eqlabel{OscPhivacResp} 
\end{gather}%
where $\Phi_{\text{cl}} $ was defined by \Eq{Fcl}.

Of crucial importance is that (\ref{eq:OscSubstEta}) 
is a genuine {\em change of functional variables\/} 
in the characteristic functional. 
We show this by solving for the inverse substitution. 
On taking the frequency-positive 
and negative parts of \Eqs{OscSubstEta} we have 
\begin{align} 
\begin{aligned}
i\eta^{(+)} (t) & = \eta^{(+)} _+(t)-\eta^{(+)} _-(t), \ \  
\sigma^{(+)}(t) = \hbar \eta^{(+)} _+(t) , \\ 
i\eta^{(-)} (t) & = \eta^{(-)} _+(t)-\eta^{(-)} _-(t), \ \  
\sigma^{(-)}(t) = \hbar \eta^{(-)} _-(t) .  
\end{aligned}%
\eqlabel{SubstEtaPM} 
\end{align}%
Solving for the frequency-positive 
and negative parts of $\eta _{\pm}(t)$ yields, 
\begin{align} 
\begin{aligned}
\eta^{(+)} _+(t) & = \frac{1}{\hbar} \sigma^{(+)}(t), \ \ 
\eta^{(-)} _+(t) = i\eta^{(-)} (t) + \frac{1}{\hbar} \sigma^{(-)}(t), \\   
\eta^{(-)} _-(t) & = \frac{1}{\hbar} \sigma^{(-)}(t), \ \ 
\eta^{(+)} _-(t) = - i\eta^{(+)} (t) + \frac{1}{\hbar} \sigma^{(+)}(t).   
\end{aligned}%
\eqlabel{SubstSigmaPM} 
\end{align}%
By using \Eq{FPNSum}, the inverse of (\ref{eq:OscSubstEta}) is found to be 
\begin{align} 
\begin{aligned} 
\eta_+(t) & = i\eta^{(-)} (t) + \frac{1}{\hbar} \sigma(t), \\    
\eta_-(t) & = - i\eta^{(+)} (t) + \frac{1}{\hbar} \sigma(t) ,   
\end{aligned}%
\eqlabel{OscSubstSigma} 
\end{align}%
proving that 
$\eta _+(t), \eta _- (t)\leftrightarrow \eta(t) , \sigma(t) $ is indeed 
a genuine functional substitution.

Substitutions (\ref{eq:OscSubstEta}), (\ref{eq:OscSubstSigma}) 
are inherently
complex. If we choose $\eta _{\pm}(t)$ to be real, 
then $\eta(t) ,\sigma(t) $ will be complex, 
and {\em vice versa\/}, but we cannot make all four functional arguments 
real simultaneously. 
It makes obvious physical sense to keep 
$\sigma(t) $ real, while for the other three it is of 
little consequence. 
A convenient formal choice is $\eta _{-}= \eta _{+}^*$, 
this making $\eta ,\sigma $ real and also turning
$\Phi (\eta _-,\eta _+)$ into a real quantity, cf.\ \Eq{PhiDef}. 
\section{Response of free bosonic fields}\chlabel{ch:LField} 
\subsection{Neutral bosons}\chlabel{app:LFN}
\subsubsection{Definitions}\chlabel{ch:BosDefN}
\newcommand{\fo}[3]{\hat #1_{\mu #3}^{#2}(\bm{r}#3,t #3)} 
In this section we extend the treatment of the harmonic oscillator to general noninteracting bosonic fields. We restrict our attention to the key point: derivation of field versions of the {\em response substitution\/} (\ref{eq:OscSubstEta}), transforming the quadratic form of contractions (\ref{eq:OscPhiVac}) into the classical expression (\ref{eq:OscPhivacResp0}).  
Here, we consider neutral bosonic fields. Charged fields will be the subject of the next section. 

To be specific, we consider a Hermitian 4-vector field described by the free-field operator $\hat Q_\mu (\bm{r},t)$. This choice suggests quantum electrodynamics. However, all our manipulations apply to the time variable and do not touch field ``labels.'' In particular, we do not make any use of the fact that $\mu $ is a 4-vector index. 
Below, $\mu $ may be regarded as an arbitrary index, e.g., a mode index running over a given set of modes. This allows all formulae to be easily adjusted to the case of an arbitrary bosonic field.

Properties of the free-field bosonic operators are considered in any text book on the quantum field theory, see, e.g., \cite{Bogol}. 
The standard quantisation approach to the bosons is based on replacing the c-number amplitudes in the mode expansion of the classical field by q-numbers: 
\eqA{ 
\hat Q_\mu (\bm{r},t) = \hbar ^{1/2}\sum_{k}
\Sbracket{\Big}{
\hat a_k \textrm{e}^{-i\omega _kt}Q^k_\mu (\bm{r}) + 
\hat a\dg_k \textrm{e}^{i\omega _kt}Q^{k*}_\mu (\bm{r}) 
}   .
}{
\eqlabel{QDef} 
}%
Here, $Q^k_\mu (\bm{r})$ are mode functions, $\omega _k$ are mode frequencies, and $\hat a\dg_k, \hat a_k$ are the standard mode creation and annihilation operators, 
\eqA{ 
\Sbracket{\big}{
\hat a_k, \hat a\dg_{k'}
} = \delta_{kk'}. 
}{
\eqlabel{ModeExpN} 
}%
We do not consider normalisation conditions for modes, allowing us to include all frequency-dependent coefficients in (\ref{eq:ModeExpN}) into the mode functions. 
The overall factor $\hbar^{1/2}$ in \Eq{QDef} allows for a convenient classical limit where $\hbar^{1/2}\hat a_k$ is replaced by the classical amplitude. It also makes the mode functions essentially classical quantities (e.g., defined by classical boundary conditions). 

With generalisation to interacting fields in mind, we wish to eliminate explicit mode expansions from considerations. To this end, we now summarise the properties of the free-field operators essential for our purposes.  
The free-field operator is naturally split into the sum of its frequency-positive and negative parts,  
\eqG{ 
\hat Q_\mu (\bm{r},t) = \hat Q^{(+)}_\mu (\bm{r},t) 
+ \hat Q^{(-)}_\mu (\bm{r},t),  \\ 
\hat Q^{(-)}_\mu (\bm{r},t) = 
\Sbracket{\Big}{
\hat Q^{(+)}_\mu (\bm{r},t)
} \dg .  
}{
\eqlabel{FieldQpm} 
}%
The frequency-positive and negative parts always apply to the time variable. Obviously, 
\eqG{ 
\hat Q^{(+)}_\mu (\bm{r},t) = \hbar ^{1/2}\sum_{k}
\hat a_k \textrm{e}^{-i\omega _kt}Q^k_\mu (\bm{r}) , 
\\ 
\hat Q^{(-)}_\mu (\bm{r},t) = \hbar ^{1/2}\sum_{k}
\hat a\dg_k \textrm{e}^{i\omega _kt}Q^{k*}_\mu (\bm{r}) .
}{
}%
Their commutator is a c-number,  
\eqMW{ 
\big[
\hat Q^{(+)}_\mu (\bm{r},t),\hat Q^{(-)}_{\mu'} (\bm{r}',t')
\big] \\ = 
\hbar \sum_k \textrm{e}^{-i\omega _k(t-t')} 
Q^k_\mu (\bm{r})Q^{k*}_\mu (\bm{r}') \equiv 
i\hbar D_{\mu \mu '}(\bm{r},\bm{r}',t-t')
,  
}{
\eqlabel{FieldBaseComm} 
}%
while two $\hat Q^{(+)}$s and two $\hat Q^{(-)}$s commute: 
\eqA{ 
\big[
\hat Q^{(+)}_\mu (\bm{r},t),\hat Q^{(+)}_{\mu'} (\bm{r}',t')
\big] = 
\big[
\hat Q^{(-)}_\mu (\bm{r},t),\hat Q^{(-)}_{\mu'} (\bm{r}',t')
\big] = 
0
,   
}{
\eqlabel{FieldBaseZeroComm} 
}%
cf.\ \Eq{OscHeis} for the oscillator. 
As seen from \Eq{FieldBaseComm}, $D_{\mu \mu '}(\bm{r},\bm{r}',t)$ is frequency-positive in respect of its time argument, 
\begin{align} 
\begin{aligned}
D^{(-)}_{\mu \mu '}(\bm{r},\bm{r}',t) = 0. 
\end{aligned}%
\end{align}%
The normal ordering is defined by placing all $\hat Q^{(-)}$s 
on the left of all $\hat Q^{(+)}$s: 
\begin{multline} %
\preprintmargin
\bm{:}\hat Q^{(+)}_\mu (\bm{r},t)\hat Q^{(-)}_{\mu'} (\bm{r}',t')\bm{:} = 
\bm{:}\hat Q^{(-)}_{\mu'} (\bm{r}',t')\hat Q^{(+)}_\mu (\bm{r},t)\bm{:}  \\ = 
\hat Q^{(-)}_{\mu'} (\bm{r}',t')\hat Q^{(+)}_\mu (\bm{r},t) 
, 
\preprintmargin
\end{multline}%
and similarly for  a larger number of factors. 
\subsubsection{Algebra of free-field Green's functions}\chlabel{ch:BosGrN}
We introduce the three contractions 
in full analogy with the 
oscillator case: 
\renewcommand{\sP}{\hspace{-0.29\columnwidth}} 
\eqA{ 
i\hbar D_{\text{F}\mu \mu '}(\bm{r},\bm{r}',t-t') \A = 
 T_+\hat Q_\mu (\bm{r},t)\hat Q_{\mu'} (\bm{r}',t') - 
\bm{:} \hat Q_\mu (\bm{r},t)\hat Q_{\mu'} (\bm{r}',t') \bm{:} ,  \\ 
- i\hbar   D^*_{\text{F}\mu \mu '}(\bm{r},\bm{r}',t-t') \A  = 
 T_-\hat Q_\mu (\bm{r},t)\hat Q_{\mu'} (\bm{r}',t') -  
\bm{:} \hat Q_\mu (\bm{r},t)\hat Q_{\mu'} (\bm{r}',t') \bm{:} ,  \\
i\hbar   D_{\mu \mu '}(\bm{r},\bm{r}',t-t') \A = 
 \hat Q_\mu (\bm{r},t)\hat Q_{\mu'} (\bm{r}',t') -  
\bm{:} \hat Q_\mu (\bm{r},t)\hat Q_{\mu'} (\bm{r}',t') \bm{:}  
. 
}{
\eqlabel{FieldContrDef} 
}%
Obviously $D_{\mu \mu '}(\bm{r},\bm{r}',t)$ here 
is the same as in \Eq{FieldBaseComm}. 
It is also easy to check that 
\eqA{ 
D_{\text{F}\mu \mu '}(\bm{r},\bm{r}',t) =  
\theta(t) D_{\mu \mu '}(\bm{r},\bm{r}',t) 
+  
\theta(-t) D_{\mu' \mu }(\bm{r}',\bm{r},-t) 
. 
}{
}%
Similarly to \Eq{OscDretbyDDF}
we define the retarded combination of contractions 
\begin{multline} %
\preprintmargin
D_{\text{R}\mu \mu '}(\bm{r},\bm{r}',t) = 
D_{\text{F}\mu \mu '}(\bm{r},\bm{r}',t) 
- D_{\mu '\mu}(\bm{r}',\bm{r},-t)  \\ = 
\theta(t) \big[D_{\mu \mu '}(\bm{r},\bm{r}',t) -  
D_{\mu' \mu }(\bm{r}',\bm{r},-t) \big]. 
\eqlabel{FieldDRDef} %
\preprintmargin
\end{multline}%
The physical meaning of this quantity will be discussed in detail in the next section. 
By making use of  (\ref{eq:FieldDRDef}) we recover the field counterpart of \Eq{DDTDRDRT}: 
\eqA{ 
D_{\mu \mu '}(\bm{r},\bm{r}',t) 
- D_{\mu' \mu }(\bm{r}',\bm{r},-t) = 
D_{\text{R}\mu \mu '}(\bm{r},\bm{r}',t) 
- D_{\text{R}\mu' \mu }(\bm{r}',\bm{r},-t) . 
}{
\eqlabel{FieldDDByDRDR} 
}%
We now recall that $D_{\mu \mu '}(\bm{r},\bm{r}',t)$ is purely frequency-%
positive and hence $D_{\mu' \mu }(\bm{r}',\bm{r},-t)$ is purely 
frequency-negative. By taking the frequency-%
positive part of (\ref{eq:FieldDDByDRDR})  we find 
\eqA{ 
D_{\mu \mu '}(\bm{r},\bm{r}',t) 
=  D_{\text{R}\mu \mu '}^{(+)}(\bm{r},\bm{r}',t) 
- D_{\text{R}\mu' \mu }^{(-)}(\bm{r}',\bm{r},-t) .  
}{
\eqlabel{FieldDByDRpm} 
}%
By using equation (\ref{eq:FieldDRDef}) 
to express $D_{\text{F}} $ we then obtain 
\eqA{ 
D_{\text{F}\mu \mu '}(\bm{r},\bm{r}',t) =  
D_{\text{R}\mu \mu '}^{(+)}(\bm{r},\bm{r}',t) + 
D_{\text{R}\mu' \mu}^{(+)}(\bm{r}',\bm{r},-t) 
.   
}{
\eqlabel{FieldDFByDRpm} 
}%
Equations (\ref{eq:FieldDByDRpm}) and (\ref{eq:FieldDFByDRpm}) generalise (\ref{eq:OscDbyDR}) and (\ref{eq:OscDFbyDR}) to the case of a neutral bosonic field. 
We see that the mode expansion enters only through equation (\ref{eq:FieldBaseComm}) for the D-function. If we assume the D-function, 
or which is the same, the field commutator, known, the concept of mode is eliminated altogether. 
\subsubsection{Response substitution for free bosonic fields}\chlabel{ch:BosRSub}
The key result on which the response reformulation of the oscillator is based is the transformation of the quadratic form of contractions into the classical-emission form, cf.\ \Eqs{OscPhiVac} and (\ref{eq:OscPhivacResp0}). For the field case, this means representing the quadratic form of the field  contractions (cf.\ \Eq{OscPhiVac})
\eqMW{ 
\ln\Phi_{\text{vac}}  (\eta _+,\eta _-) \\ = {i\hbar}
 \sum _{\mu \mu '}\int dt dt' d^3\bm{r}d^3\bm{r}' 
\Sbracket{\bigg}{
- \frac{1}{2}\eta _+^{\mu }(\bm{r},t)\eta _+^{\mu '}(\bm{r}',t')
D_{\text{F}\mu \mu '}(\bm{r},\bm{r}',t-t')
\\ + \frac{1}{2}\eta _-^{\mu }(\bm{r},t)\eta _-^{\mu '}(\bm{r}',t')
D^*_{\text{F}\mu \mu '}(\bm{r},\bm{r}',t-t')
+ \eta _-^{\mu }(\bm{r},t)\eta _+^{\mu '}(\bm{r}',t')
D_{\mu \mu '}(\bm{r},\bm{r}',t-t')
} , 
}{
\eqlabel{FieldCForm} 
}%
as 
\begin{multline} %
\preprintmargin
\ln\Phi_{\text{vac}}  (\eta _+,\eta _-) \\ = 
\sum _{\mu \mu '}\int dt dt' d^3\bm{r}d^3\bm{r}' 
\eta ^{\mu }(\bm{r},t) 
D_{\text{R}\mu \mu '}(\bm{r},\bm{r}',t-t')   
\sigma ^{\mu '}(\bm{r}',t') , 
\eqlabel{FieldDRForm} 
\preprintmargin
\end{multline}%
where $D_{\text{R}\mu \mu '}(\bm{r},\bm{r}',t-t')$ is the retarded combination of contractions defined by (\ref{eq:FieldDRDef}). 
In omitting integration limits in (\ref{eq:FieldCForm}) and (\ref{eq:FieldDRForm}) we follow endnote \cite{IntRange}. 
The relation between $\eta _{\pm}$ and $\eta ,\sigma $ is found by substituting \Eqs{FieldDByDRpm} and (\ref{eq:FieldDFByDRpm}) into (\ref{eq:FieldCForm}) and manipulating the emerging expression in detailed analogy with \Eqs{OscPropPP0}--(\ref{eq:OscPropPM}). As a result we indeed arrive at \Eq{FieldDRForm} with 
\begin{gather} 
\begin{aligned}
\eta^{\mu }(\bm{r},t) & = -i \big[ 
\eta^{\mu }_+(\bm{r},t)- \eta^{\mu }_-(\bm{r},t)\big], \\ 
\sigma^{\mu }(\bm{r},t) & = \hbar \big[ 
\eta^{\mu (+)}_+(\bm{r},t)+\eta^{\mu (-)}_-(\bm{r},t)\big] . 
\end{aligned}%
\eqlabel{SubstEta} 
\end{gather}%
This is a field counterpart of the substitution (\ref{eq:OscSubstEta}). It is now clear that any relation for the oscillator may be generalised to the field case by simply adding the field ``labels'' to the time argument, $t \to \mu ,\bm{r},t$. 
It may be easily checked that this indeed generalises {\em all\/} relations derived in the previous section to neutral bosonic fields. We also note that substitutions (\ref{eq:SubstEta}) bear no trace of the mode expansion (\ref{eq:QDef}) which we used as a starting point. 
\subsection{Charged bosons}\chlabel{app:LFCh}
\subsubsection{Definitions}
Referring the reader for details to any textbook on the quantum field theory, here we only list the key properties of the free-field operators. A charged bosonic field is described by a conjugated pair of field operators, \begin{align} 
\begin{aligned}
\fo{F}{}{} & = \fo{A}{(+)}{} + \fo{B}{(-)}{}, \\ 
\fo{F}{\dag}{} & = \fo{A}{(-)}{} + \fo{B}{(+)}{}
. 
\end{aligned}%
\end{align}%
We use distinct notation for the frequency-positive and negative 
parts of the field operator to emphasise their distinct physical nature: 
$\fo{F}{(+)}{} = \fo{A}{(+)}{}$ annihilates particles while 
$\fo{F}{(-)}{} = \fo{B}{(-)}{}$ creates antiparticles. Similarly, 
$\fo{F}{\dag(+)}{} = \fo{B}{(+)}{}$ annihilates antiparticles while 
$\fo{F}{\dag(-)}{} = \fo{A}{(-)}{}$ creates particles. 
The operators $\hat A^{(+)}$, $\hat A^{(-)}$, $\hat B^{(+)}$, 
and $\hat B^{(-)}$ 
pairwise commute except the two pairs: 
\begin{align} 
\begin{aligned}
\big[\fo{A}{(+)}{},\fo{A}{(-)}{'}\big] & \equiv 
i\hbar D^A_{\mu \mu '}(\bm{r},\bm{r}',t-t'), 
\\ 
\big[\fo{B}{(+)}{},\fo{B}{(-)}{'}\big] & \equiv 
i\hbar D^B_{\mu' \mu}(\bm{r}',\bm{r},t'-t) . 
\end{aligned}%
\eqlabel{ChargedDDef} 
\end{align}%
Function $D^A$ is frequency-positive while $D^B$ is 
frequency-negative, 
\begin{align} 
\begin{aligned}
D^{A(+)}_{\mu \mu '}(\bm{r},\bm{r}',t) = 
D^{B(+)}_{\mu \mu '}(\bm{r},\bm{r}',t) = 
0.   
\end{aligned}%
\eqlabel{ChargedDFPN} 
\end{align}%
\subsubsection{Algebra of Green's functions}
From now on, for brevity we drop the field labels; similarly to the neutral case they can always be restored by replacing $t \to \mu ,\bm{r},t$. 
Unlike in the neutral case we have now four contractions, 
\begin{gather} 
\begin{aligned}
T_+ \hat F(t) \hat F \dg (t') -\, \bm{:}\hat F(t) \hat F \dg (t')\bm{:}\, & =  i\hbar D_{\text{F}}(t-t') ,  
\\ 
T_- \hat F(t) \hat F \dg (t') -\, \bm{:}\hat F(t) \hat F \dg (t')\bm{:}\, & =  -i\hbar D\dg_{\text{F}}(t-t') ,  
\\ 
\hat F(t) \hat F \dg (t') -\, \bm{:}\hat F(t) \hat F \dg (t')\bm{:}\, & =  i\hbar D^A(t-t') ,  
\\ 
\hat F \dg(t') \hat F (t) -\, \bm{:}\hat F(t) \hat F \dg (t')\bm{:}\, & =  i\hbar D^B(t-t') ,  
\end{aligned}%
\eqlabel{ChargedContr} 
\end{gather}%
where the normal operator ordering is defined in the usual way as  
\eqA{ 
\bm{:}\hat A^{(+)}(t)\hat A^{(-)}(t')\bm{:}\, = \,
\bm{:}\hat A^{(-)}(t')\hat A^{(+)}(t)\bm{:} \, = 
\hat A^{(-)}(t')\hat A^{(+)}(t) ,  
}{
}%
and similarly for the $B$s. 
Hermitian conjugation of c-number kernels is defined in the standard way, 
\begin{align} 
\begin{aligned}
D\dg _{\text{F}}(t) = D^* _{\text{F}}(-t) .    
\end{aligned}%
\eqlabel{HermKern} 
\end{align}%
This formula implies that we regard $D _{\text{F}}$ as a c-number operator kernel $D _{\text{F}}(t-t')$, and similarly for other kernels. 
The second line in (\ref{eq:ChargedContr}) is Hermitian conjugation 
of the first one. 
The other two expressions are independent; 
they are Hermitian so that $D^A$ and $D^B$ are anti-Hermitian: 
\begin{align} 
\begin{aligned}
D^{A\dag}(t) = - D^A(t) , \ \ D^{B\dag}(t) = - D^B (t). 
\end{aligned}%
\eqlabel{DAHerm} 
\end{align}%
The time and normal orderings apply formally to 
commuting pairs as well but in fact do nothing, e.g. 
\eqA{ 
T_+ \hat F(t) \hat F (t') \A =   
\bm{:}\hat F(t) \hat F (t')\bm{:} = \hat F (t) \hat F(t') , 
}{
}%
etc. 
The corresponding contractions are zero. 

It should not be overlooked that $D_{\text{F}}(t)$ is now not symmetric 
with respect to time reversal. 
From (\ref{eq:ChargedDDef}) and (\ref{eq:ChargedContr}), 
\begin{align} 
\begin{aligned}
D_{\text{F}}(t) & = \theta(t) D^A(t) + \theta(-t) D^B(t), \\
D\dg_{\text{F}}(t) & = \theta(-t) D^{A\dag}(t) + \theta(t) D^{B\dag}(t), \\ 
\end{aligned}%
\end{align}%
where 
the second line is conjugation of the first, cf.\ (\ref{eq:HermKern}). 
It may therefore seem that one can construct two
independent retarded combinations: one by subtracting $D^B$ 
from $D_{\text{F}} $ and the other---by subtracting $D^{A\dag}$ 
from $D\dg_{\text{F}} $. 
These turn out to be 
two ways of expressing the same function 
\begin{align} 
\begin{aligned}
D_{\text{R}}(t) & = D_{\text{F}}(t) -  D^B(t) 
= \theta(t) \big [ D^A(t) - D^B(t)\big ], \\
D_{\text{R}}(t) & = D^{\dag}_{\text{F}}(t) - D^{A\dag}(t) = 
\theta(t) \big [D^{B\dag}(t) -  D^{A\dag}(t)\big ].  
\end{aligned}%
\end{align}%
Equivalence of these two definitions 
is clearly seen from (\ref{eq:DAHerm}). 
We note that $D\R$ is now complex so that 
one can indeed say that we have two {\em real\/} 
retarded combinations, but this does not appear to have any physical implications. 
Having a complex retarded Green's function is a natural property of 
a charged field which is a quantised equivalent of a complex 
c-number field. 

Combining the definitions of $D\R$ we find the ``charged'' 
equivalent of (\ref{eq:DDTDRDRT}) 
\begin{align} 
\begin{aligned}
D\R(t) - D\R\dg(t) = D^A(t) - D^B(t) .  
\end{aligned}%
\end{align}%
By using (\ref{eq:ChargedDFPN}) we find 
\begin{align} 
\begin{aligned}
D^A(t) & = D\R^{(+)}(t) -  D\R\dgp{(+)}(t), \\ 
D^B(t) & =  D\R\dgp{(-)}(t) - D\R^{(-)}(t),   
\end{aligned}%
\eqlabel{ChargedD} 
\end{align}%
while by reusing the definitions of $D\R$ to express 
$D_{\text{F}} $ and $D\dg_{\text{F}} $ we have  
\begin{align} 
\begin{aligned}
D_{\text{F}} (t) & = D\R(t) + D^B(t) = D\R^{(+)}(t) +  D\R\dgp{(-)}(t), 
\\  
D\dg_{\text{F}} (t) & =  D\R(t) + D^{A\dag}(t) 
= D\R\dgp{(+)}(t) +  D\R^{(-)}(t). 
\end{aligned}%
\eqlabel{ChargedDF} 
\end{align}%
In obtaining the second equation here we used that 
\begin{align} 
\begin{aligned}
D\R\dgp{(\pm)}(t) = D\R^{(\pm)\dag}(t).   
\end{aligned}%
\end{align}%
This holds for any kernel; it also 
makes it evident that the two equations in (\ref{eq:ChargedDF}) 
are indeed conjugate. 
\subsubsection{The response substitution}
By making use of \Eqs{ChargedD} and (\ref{eq:ChargedDF}), one can identify the response substitution for charged fields. 
In full notation the ``charged'' version of \Eq{FieldDRForm} is written as 
(cf.\ endnote \cite{IntRange}) 
\begin{multline} 
i\hbar \sum _{\mu \mu '}\int dt dt' d^3\bm{r}d^3\bm{r}' \Big[
- \bar \eta _+^{\mu }(\bm{r},t)\eta _+^{\mu '}(\bm{r}',t')
D_{\text{F}\mu \mu '} (\bm{r},\bm{r}',t-t')
\\ + 
\bar \eta _-^{\mu }(\bm{r},t)\eta _-^{\mu '}(\bm{r}',t')
D_{\text{F}\mu' \mu }^* (\bm{r}',\bm{r},t'-t)
\\ + 
\bar \eta _-^{\mu }(\bm{r},t)\eta _+^{\mu '}(\bm{r}',t')
D^A_{\mu \mu '} (\bm{r},\bm{r}',t-t')
+ \bar \eta _+^{\mu }(\bm{r},t)\eta _-^{\mu '}(\bm{r}',t')
D^B_{\mu \mu '} (\bm{r},\bm{r}',t-t')
\Big ]
\\ = 
\sum _{\mu \mu '}\int dt dt' d^3\bm{r}d^3\bm{r}' \Big[
\bar \eta ^{\mu }(\bm{r},t)
D_{\text{R}\mu \mu '} (\bm{r},\bm{r}',t-t')
\sigma ^{\mu' }(\bm{r}',t') 
+ 
\eta ^{\mu }(\bm{r},t)
D^*_{\text{R}\mu \mu '} (\bm{r},\bm{r}',t-t')
\bar \sigma ^{\mu' }(\bm{r}',t') \Big]
, 
\eqlabel{ChDRForm} 
\end{multline}%
where 
\begin{gather} 
\begin{aligned}
\eta^{\mu }(\bm{r},t) & = -i \big[ 
\eta^{\mu }_+(\bm{r},t)- \eta^{\mu }_-(\bm{r},t)\big], \\ 
\sigma^{\mu }(\bm{r},t) & = \hbar \big[ 
\eta^{\mu (+)}_+(\bm{r},t)+\eta^{\mu (-)}_-(\bm{r},t)\big] , \\ 
\bar \eta^{\mu }(\bm{r},t) & = -i \big[ 
\bar \eta^{\mu }_+(\bm{r},t)- \bar \eta^{\mu }_-(\bm{r},t)\big], \\ 
\bar \sigma^{\mu }(\bm{r},t) & = \hbar \big[ 
\bar \eta^{\mu (+)}_+(\bm{r},t)+\bar \eta^{\mu (-)}_-(\bm{r},t)\big]  . 
\end{aligned}%
\eqlabel{ChSubstEta} 
\end{gather}%
Equations (\ref{eq:ChSubstEta}) are simply the substitution (\ref{eq:SubstEta}) repeated twice. 
\section{Results and discussion}\chlabel{ch:Disc}
\subsection{Response reformulation of bosons}%
\chlabel{ch:RespRef} 
Having proven that the results for the harmonic oscillator are trivially generalised to arbitrary noninteracting bosonic fields, for simplicity we return in the ensuing discussion to the oscillator case. 
From equations (\ref{eq:FFF}) and 
(\ref{eq:OscPhivacResp}) we have  
\eqA{ 
\Phi (\eta _-,\eta _+;j) = \Phi_{\text{cl}}  (\eta;j+\sigma )
\Phi_{\text{in}}  (\eta) 
\equiv \Phi \R(\eta ;j+\sigma ) .  
}{
\eqlabel{FOscFin0} %
}%
We see that the variable $\sigma $ is redundant: all quantum dynamical and statistical properties of the oscillator emerge as its self-radiation and response properties through the functional 
\begin{multline} %
\preprintmargin
\Phi \R(\eta ;j) = \Phi_{\text{cl}}  (\eta;j )
\Phi_{\text{in}}  (\eta) \\ = \nav{
\exp\int dt\, \eta (t)\left[
\hat q (t) + \int dt' D_R(t-t')
j(t') 
\right] 
}
. 
\eqlabel{PhiRJ} 
\preprintmargin
\end{multline}%
Describing the system in terms of the functional $\Phi \R(\eta ;j)$ will be called 
a {\em response formulation\/}. 
This functional can be re-expressed in two equivalent forms. 
Firstly, by noting that the quantity in square brackets equals 
$\hat q(t) + q_j(t) = \hat q_j(t)$, cf.\ \Eqs{qjbyq} and (\ref{eq:qj}), we can write 
\begin{gather} 
\begin{aligned}
\Phi \R(\eta ;j) = \nav{
\exp\int dt\, \eta (t)
\hat q_j (t) 
}. 
\end{aligned}%
\eqlabel{qjnorm} 
\end{gather}%
That is, $\Phi \R(\eta ;j)$ is a characteristic functional of the normal averages of operator $\hat q_j (t)$. 
Equation (\ref{eq:qjnorm}) thus establishes a connection between the response formulation and Glauber's macroscopic photodetection theory \cite{GlauberPhDet,KellyKleiner,GlauberTN,MandelWolf}. 
Suitably generalised, this relation will play a fundamental role in the analyses of interacting systems. 
Secondly, introducing the P-distribution $P(\alpha)$ characteristic of the initial state, 
\begin{gather} 
\begin{aligned}
\nav{\hat a\dgp{n} \hat a^m} = \int \frac{d^2\alpha}{\pi } P(\alpha)\, 
\alpha ^{*n} \alpha ^{m} \equiv \av{\alpha ^{*n} \alpha ^{m}},
\end{aligned}%
\end{gather}%
we can rewrite (\ref{eq:PhiRJ}) as 
\eqA{ 
\Phi \R(\eta ;j) = 
\av{
\exp\int dt\, \eta (t)\left[
q_{\text{in}} (t) + \int dt' D_R(t-t')
j(t')\right] }
, 
}{
\eqlabel{FOscFin} 
}%
where 
\begin{align} 
\begin{aligned}
q_{\text{in}} (t) = \frac{q_0}{ \sqrt{2}}\left(
\alpha \textrm{e}^{-i\omega_0 t} + \alpha^* \textrm{e}^{i\omega_0 t}
\right) 
.   
\end{aligned}%
\eqlabel{OscQinDef} 
\end{align}%
We use the bar as a notation for the averaging over the initial P-distribution to emphasise the analogy between the response formulation of the quantum oscillator and a c-number description of the classical oscillator. 
Were we to give this symbol back its customary meaning of a classical statistical averaging, (\ref{eq:OscQinDef}) would become a characteristic functional of multi-time momenta of the c-number oscillator coordinate $q(t)$:  
\begin{gather} 
\begin{gathered} 
q(t) = q_{\text{in}} (t) + \int dt' D_R(t-t')
j(t'), 
\\
\av{q(t_1)\cdots q(t_m)} = \left . \frac{\delta^m \Phi \R(\eta ;j)}
{\eta (t_1)\cdots\eta (t_m)} \right | _{\eta =0}. 
\end{gathered}%
\eqlabel{PhiClOsc} 
\end{gather}%
Except for the possible nonpositivity of $P(\alpha)$, \Eqs{FOscFin} and (\ref{eq:PhiClOsc}) appear fully --- and astonishingly --- classical.

Using $\eta (t),\sigma (t)$ as arguments in the 
charactristic functional brings out perfect physical sense. 
Function $\sigma (t)$ has turned out to be combined with the external 
c-number driving force, a property characterising the 
{\em input\/} of a system; we shall in future see that this property is quite general and also holds in the case of nonlinear interacting fields. 
Setting $\sigma (t)$ to zero turns $\Phi(\eta _-,\eta _+;j)$ 
into a characteric functional for the normally ordered 
averages; again, this is quite natural a 
property for the {\em output\/} of a system, recalling that the 
normally ordered operator averages emerge in the macroscopic detection 
theory of Glauber \cite{GlauberPhDet}. 
We see also that, whether the initial state of the quantum 
oscillator is quantum or classical, 
its quantum dynamics, i.e., 
everything related to the response to the driving force 
as well as vacuum fluctuations,  
emerges in fully classical form. 
\subsection{Kubo's linear response as a quantisation 
condition}\chlabel{ch:Kubo} 
Equation (\ref{eq:qjnorm}) makes evident the connection between the response formulation and Glauber's photodetection theory. 
Our immediate goal is to clarify another important connection, namely, between the response formulation and Kubo's linear response theory \cite{Kubo}. 
The way $D\R$ was introduced into the quantum treatment in section \ref{ch:OscAlg} 
gives a misleading impression of pure luck: 
we just happened to notice that a particular combination of the quantum Green's functions coincides with the retaded Green's function of the classical oscillator. In fact the situation is far more physical. 
By using the primary definitions of $D_{\text{F}} $ and $D$ as contractions, 
cf.\ \Eqs{DFDef}--(\ref{eq:DDef}), we rewrite \Eq{OscDretbyDDF} as   
\eqA{ 
D\R(t) = -\frac{i}{\hbar } \left[
T_+ \hat q(t) \hat q(t') - \hat q(t') \hat q(t)
\right] = 
-\frac{i}{\hbar } \theta(t-t') \qav{
\left[
\hat q(t), \hat q(t')
\right]
}. 
}{
\eqlabel{DRResp} %
}%
The commutator on the RHS here is a c-number, so setting it formally 
under the quantum averaging does not change anything. 
We introduced the averaging to make it obvious that 
the RHS of (\ref{eq:DRResp}) 
is in fact Kubo's formula for the 
linear response function of the quantum oscillator 
with respect to the external 
current in Hamiltonian (\ref{eq:HamJ}).  
Thus the actual physical message of equation (\ref{eq:OscDretbyDDF}) is that the {\em linear responses of the classical and quantum 
oscillators coincide\/}.
In other words, if we restrict the measurement to the linear response, the quantum and the classical oscillators become {\em experimentally indistinguishable\/}.   

Statements like this one are at the core of the very concept of 
{\em quantisation\/}. What it says, in essence, is, ``the change 
from the c-number to the q-number description preserves certain 
{\em physical\/} aspects of the situation.'' The standard 
quantisation approach for the oscillator says, ``retain the 
equations of motion for the coordinate and momentum while making these 
quantities non-commuting.'' Unlike equations 
of motion, response is an observable property so specifying it 
rather than the equations 
looks as much more physical an approach. 
We now show that a consistent 
quantisation approach to the oscillator can indeed be devised based on 
\Eq{DRResp}. 
To start with, we show that 
this equation allows one to express the commutator 
by $D\R$. Firstly, we rewrite it as 
\begin{gather} 
\theta(t-t')
\big[\hat q (t),\hat q (t')\big] = 
i\hbar D_{\text{R}}(t-t') . %
\eqlabel{KuboFree1} 
\end{gather}%
Swapping here $t \leftrightarrow t' $ 
and then using the antisymmetry of 
the commutator yields: 
\eqA{ 
\theta(t'-t)
\big[\hat q (t'),\hat q (t)\big] =  
- \theta(t'-t)
\big[\hat q (t),\hat q (t')\big] =  
i\hbar D_{\text{R} }(t'-t) . 
}{
\eqlabel{KuboFree2} 
}%
Subtracting (\ref{eq:KuboFree2}) from (\ref{eq:KuboFree1}) we find 
\begin{gather} 
\big[\hat q (t),\hat q (t')\big] =  
i\hbar \big[D_{\text{R}}(t-t') 
- D_{\text{R}}(t'-t)\big] . 
\eqlabel{CommByKubo} 
\end{gather}%
We see that, for the free field, the information contained in 
the two-time commutator is the linear response of the field. 
Paradoxical as it may sound, 
the noncommutivity of field operators, 
this quintessence of the quantum, for the oscillator is simply a technical 
way of describing response. 
This statement is fully supported 
by the formal treatment and must therefore be taken seriously. 
We shall see in the future that the technical equivalence of noncommutivity and response is true also for interacting nonlinear 
systems.

Full set of observables for the oscillator includes momentum $\hat p(t)$. 
Commutators involving $\hat p(t)$ are found on  
supplementing (\ref{eq:CommByKubo}) by 
\begin{gather} 
\begin{aligned}
\hat p(t)= m\, \dot {\hat q}(t) . 
\end{aligned}%
\end{gather}%
In particular, by making use of \Eq{OscDRExpl},  
\eqA{ 
\big[\hat q (t),\hat p (t')\big] =  
i\hbar m\frac{d}{dt'} \big[D_{\text{R}}(t-t') 
- D_{\text{R}}(t'-t)\big] = i\hbar \cos \omega _0 (t-t')
. 
}{
}%
The canonical quantisation condition follows: 
\begin{gather} 
\begin{aligned}
\big[\hat q (t),\hat p (t)\big] = i\hbar . 
\end{aligned}%
\end{gather}%
We note that the definition of momentum is quite general 
while all physical particulars about the oscillator enter 
through the response. 
\subsection{C-number sources in quantum mechanics: Kubo {\em versus\/} Schwinger}%
\chlabel{ch:Schwinger} 
\subsubsection{The Kubo and Schwinger currents}%
It is instructive to clarify the relation between the concepts of external c-number current in Schwinger's and Kubo's approaches. To this end, consider the evolution operator in the interaction picture. It obeys the equation  
\eqA{ 
i\hbar {\dot{{\Heis{U}}}}(t) = j(t) \hat q(t){{\Heis{U}}}(t) .   
}{
\eqlabel{UpE} 
}%
solved by 
\begin{gather} 
\begin{aligned}
{{\Heis{U}}}(t) = T_+ \exp\left[
-\frac{i}{\hbar } \int_{-\infty}^t dt' j(t') \hat q(t')
\right] . 
\end{aligned}%
\eqlabel{Uj} 
\end{gather}%
Following Schwinger \cite{SchwingerC}, we distinguish evolution forward in time from evolution back in time. The former is determined by current $j=j_+(t)$, the latter is commanded by an independent current $j=j_-(t)$. Formally, this means considering a conjugate pair of evolution operators:  \begin{gather} 
\begin{aligned}
{{\Heis{U}}}_+(t) = T_+ \exp\left[
-\frac{i}{\hbar } \int_{-\infty}^t dt' j_+(t') \hat q(t')
\right] . 
\end{aligned}%
\eqlabel{Up} 
\end{gather}%
propagates the system forward in time, while the conjugate evolution operator
\begin{gather} 
\begin{aligned}
{{\Heis{U}}}\dg_-(t) = T_- \exp\left[
\frac{i}{\hbar } \int_{-\infty}^t dt' j_-(t') \hat q(t')
\right]  
\end{aligned}%
\eqlabel{Um} 
\end{gather}%
propagates it back in time. Functional $\Phi (\eta _-,\eta _+)$ can then be written as a Schwinger-style amplitude, 
\begin{gather} 
\begin{aligned}
\Phi (\eta _-,\eta _+) = \qav{\Heis{S}_-\dg\,\Heis{S}_+},  
\end{aligned}%
\end{gather}%
where $\Heis{S}_+$ and $\Heis{S}_-\dg$ are the forward-in-time and back-in-time S-matrices, 
\begin{gather} 
\begin{aligned}
\Heis{S}_+ = \Heis{U}_+(\infty) & = T_+ \exp\left[
-\frac{i}{\hbar } \int dt' j_+(t') \hat q(t')
\right], 
\\ 
\Heis{S}_-\dg = \Heis{U}_-\dg(\infty) & = T_- \exp\left[
\frac{i}{\hbar } \int dt' j_+(t') \hat q(t')
\right] 
, 
\end{aligned}%
\end{gather}%
and 
\begin{gather} 
\begin{aligned}
j_{\pm}(t) = \hbar \eta_{\pm}(t). 
\end{aligned}%
\end{gather}%
Transformations (\ref{eq:OscSubstEta}) then read 
\begin{gather} 
\begin{aligned}
\hbar \eta(t) &= -i\big[j_+(t) - j_-(t)\big], \ \ j = j_+^{(+)}(t) + j_-^{(-)}(t). 
\end{aligned}%
\eqlabel{OSEJJ} 
\end{gather}%

We see that the dependence of the Schwinger amplitude on the $j_{\pm}$ pair contains full information on the response properties of the system, but in a fairly coded and tangled manner. If $j_+ = j_-$, the Schwinger amplitude is simply unity. We have to consider $j_+$ and  $j_-$ as independent quantities; in suffices to assume that $j_+$ is complex and $j_- = j_+^*$. There is no possibility to have $j$, $j_+$ and  $j_-$ real simultaneously. Unlike Kubo's current, Schwinger's currents are formal quantities. Physically, it is natural to consider the $\eta ,j$  variable pair as primary, and the $ j _{\pm}$ pair---as secondary. 
In other words, we regard the response viewpoint as a primary physical viewpoint. 
Description of the system in terms of its dependence on the Schwinger currents is from this perspective secondary. 

\subsubsection{Response formulation and the operator ordering}%
\newcommand{\jW}{j_{\text{W}} } 
\newcommand{\FW}{\Phi _{\text{in}}^ {\text{W}} } 
One may wonder what if we used some other way of defining the Kubo current by the Schwinger currents. For instance, let us define, 
\begin{gather} 
\begin{aligned}
\hbar \eta (t) = - i \left[
 j_+(t) -  j_-(t)
\right], \ \ 
\jW(t) = \frac{1}{2} \left[
 j_+(t) +  j_-(t)
\right]  . 
\end{aligned}%
\eqlabel{SubsW} 
\end{gather}%
Does functional $\Phi \left(
\frac{ j_-}{\hbar } , \frac{ j_+}{\hbar }
\right) $ have any meaning in variables $\eta ,\jW $? Due to {\em ad hoc\/} nature of (\ref{eq:SubsW}), it is even unclear if $j_{\text{W}}$ can be interpreted dynamically in terms of Hamiltonian (\ref{eq:HamJ}) with $j\to j_{\text{W}} $.  
{\em A priori\/}, calling $\jW$ a Kubo current is in no way justified, but this interpretation of $\jW$ indeed holds. 
Namely, in appendix \ref{app:Symm} the following relation is derived: 
\eqA{ 
\Phi (\eta _-,\eta _+) = \Phi \R^{\text{W}}(\eta ,\jW) = 
\qav{W\exp\int dt\, \eta (t) \hat q_{\jW}(t)}, 
}{
\eqlabel{WResp} %
}%
where $W\cdots$ denotes the symmetric (Weyl) ordering \cite{MandelWolf} of the creation and annihilation operators. 
This way, substitution (\ref{eq:SubsW}) introduces an alternative response viewpoint based on Kubo's Hamiltonian and Weyl's operator ordering. It can also be shown that yet another way of defining Kubo's current, 
\begin{gather} 
\begin{aligned}
j_{\text{A}}(t) = j_+^{(-)}(t)+j_-^{(+)}(t) , 
\end{aligned}%
\end{gather}%
is equivalent to combining Kubo's Hamiltonian with antinormal operator ordering, 
\eqA{ 
\Phi (\eta _-,\eta _+) = \Phi \R^{\text{A}}(\eta ,j_{\text{A}}) = 
\qav{A\exp\int dt\, \eta (t) \hat q_{j_{\text{A}}}(t)}. 
}{
}%
It is then only natural to come up with a conjecture that specifying the way the Kubo current is expressed by the Schwinger currents equals specifying an operator ordering used to define the output of the system. Physically meaningful discussion of this conjecture must be postponed till general properties of response reformulation are understood for interacting systems which is a subject of our forthcoming papers. Here we restrict ourselves to two remarks. Firstly, we note that this result is certainly counter-intuitive. Contrary to expectations, changing the way Kubo's current is expressed by Schwinger's currents affects not the fact that Kubo's current may be interpreted as such but rather the way we describe the system in phase-space terms. Redefining Kubo's current affects not the input but the output of the system. Secondly, it is the characteristic property of the normal ordering to map the quantum oscillator {\em identically\/} on the classical oscillator. All other orderings introduce various ``vacuum noises'' resulting in apparent dissimilarity of the quantum and classical viewpoints. 
\section{Conclusion}\chlabel{ch:Conc} 
We have introduced a {\em quantum-statistical response formulation\/} of noninteracting bosons relating all quantum dynamical properties of the system to its response properties. This formulation is, by construction, equivalent with the conventional description of the system in terms of Green's functions 
in Schwinger's closed-time-loop formalism. It is also shown to be naturally related to Glauber's photodetection theory and Kubo's linear response theory. Its most important property is that a physically classical system is in response formulation also formally classical, 
i.e., described by c-numbers and probabilities. In our forthcoming papers we shall generalise the response formulation to interacting quantum fields, bosonic as well as fermionic.
\section{Acknowledgments}
S.\ Stenholm wishes to thank the Institut f\"ur Quantenphysik, Universit\"at Ulm, and Prof.\ W.P.\ Schleich for generous hospitality. 

 
\appendix
\section{Wick's theorem}\flashlabel{app:Wick} 
In its initial form, Wick's theorem expresses time-ordered products of free-field operators by their normal products. 
It can easily be verified that proof of Wick's theorem (see, e.g., \cite{Bogol}) is based solely on the fact that the time axis is linearly ordered. It can therefore be extended to field operators defined on any linearly ordered set. In particular \cite{Keldysh}, it holds for the double-time-ordering which, as we have already mentioned, can alternatively be regarded as ordering on the C-contour shown in \mbox{Fig.\ \ref{fig:C}}. 

In the case of harmonic oscillator, Wick's theorem for the double-time ordering \cite{Keldysh} may conveniently be written as a closed relation, 
\begin{multline} %
\preprintmargin
T_- \exp \bigg [ i \int dt\,\eta _-(t) {\hat{{q}}}(t)\bigg ] 
\  
T_+ \exp \bigg [-i \int dt\,\eta _+(t) {\hat{{q}}}(t)\bigg ] \\ = 
\Phi_{\text{vac}} (\eta _-,\eta _+)\ \bm{:} 
\exp \int dt\, \eta (t)
{\hat{{q}}}(t)
\bm{:}
\, ,  
\eqlabel{Wick} %
\preprintmargin
\end{multline}%
where $\Phi_{\text{vac}} (\eta _-,\eta _+)$ is given by \Eq{OscPhiVac}. Equation (\ref{eq:InVac}) then follows simply by averaging (\ref{eq:Wick}). We start from proving relation (\ref{eq:Wick}) for the $T_+$-ordering, the case of double-time ordering being then a matter of straightforward generalisation. Wick's theorem for the $T_+$-ordering stipulates that 
``a time-ordered product of free-field operators equals a sum of normal 
products resulting from all possible ways of replacing pairs of operators by their contractions,'' the latter being defined by \Eq{DFDef}. 
As was first noticed by Hori \cite{Hori}, the {\em algebraic structure\/} 
corresponding to the ``sum of operator products with all possible contractions'' is produced by an exponential differential operator 
applied to a product of c-number functions $q(t_1)\cdots q(t_m)$.   
Hori's observation allows one to write Wick's theorem as 
\begin{gather} 
\begin{aligned}
T_+ \hat q(t_1)\cdots \hat q(t_m) = \bm{:} 
\left.\big[\exp({\Delta})\, q(t_1)\cdots q(t_m)
\big]\right|_{q\to \hat q} \bm{:} 
\, ,    
\end{aligned}%
\eqlabel{Hori} 
\end{gather}%
where 
\begin{gather} 
\begin{aligned}
\Delta = \frac{i\hbar }{2} \int dt dt' D_F(t-t') 
\frac{\delta^2}{\delta q(t)\delta q(t')} .  
\end{aligned}%
\end{gather}%
To prove this, we note that applying $\Delta$ results 
in correct patterns with one contraction: 
\begin{gather} 
\begin{aligned} 
\Delta q(t) &= 0 , \\
\Delta q(t_1)q(t_2) & = i\hbar D_F(t_1 - t_2) , \\
\Delta q(t_1)q(t_2)q(t_3) & = i\hbar D_F(t_1 - t_2)q(t_3)\,  + \\ 
& \hspace{-0.2\columnwidth} 
i\hbar D_F(t_1 - t_3)q(t_2) + i\hbar D_F(t_2 - t_3)q(t_1)
, \\ & \hspace{0.026\columnwidth}\vdots
\end{aligned}%
\eqlabel{OneDelta} 
\end{gather}%
All terms enter with coefficient 1 as required. 
Furthermore, $\Delta^2$ produces terms with two contractions, each  
with coefficient 2; this reflects the two ways a pair of given 
contractions can be added sequentially. 
The same reasoning shows that $\Delta^n$ produces terms of $n$ contractions with coefficients $n!$. 
The correct combination of terms with all possible contractions, 
each entering with coefficient 1, including 
the term without contractions, 
is therefore indeed produced by the sum 
\begin{gather} 
\begin{aligned}
1 + \sum_{n=1}^{\infty} \frac{1}{n!}\,\Delta^n = \exp({\Delta}). 
\end{aligned}%
\eqlabel{AllDeltas} 
\end{gather}%

By making use of (\ref{eq:Hori}) we have 
\begin{multline} %
\preprintmargin
T_+ \exp \left[
-i \int dt\, \eta _+(t) \hat q(t)
\right]  \\ = 
\bm{:} 
\left.\left\{
\exp({\Delta}) \,\exp \left[
-i \int dt\, \eta _+(t) q(t)
\right]
\right\} \right|_{q\to \hat q} \bm{:} 
\, .   
\eqlabel{WickHori} %
\preprintmargin
\end{multline}%
The quantity in curly brackets is easily calculated using that,  
for any pair of functionals $F(s )$ and $G(q)$, we have 
\begin{gather} 
\begin{gathered} 
F\left(
\frac{\delta}{\delta q} 
\right) G(q) = \left.F\left(
\frac{\delta}{\delta q'} 
\right) G(q+q')\right|_{q'=0} , \\ 
\left.F\left(
\frac{\delta}{\delta q} 
\right) G(q)\right|_{q=0} = \left.G\left(
\frac{\delta}{\delta s }
\right) F(s)\right|_{s=0}  . 
\end{gathered}%
\end{gather}%
This can be proven directly by expanding the functionals into their functional Taylor series. By making use of these relations, the calculation reduces 
to the application of a functional shift operator: 
\begin{multline} 
\exp({\Delta}) \,\exp \left[
-i \int dt\, \eta _+(t) q(t)
\right] = 
\exp \left\{
-i \int dt\,\eta _+(t) \left[ q(t) + \frac{\delta}{\delta s(t)} 
\right]\right\} \\ \times \exp \left[
\frac{i\hbar }{2} \int dt dt' D_F(t-t') 
s(t) s(t') 
\right] \bigg | _{s=0} \\ =
\exp \left[
-i \int dt\,\eta _+(t) q(t)   
\right]  \exp \left[ - 
\frac{i\hbar }{2} \int dt dt' D_F(t-t') 
\eta _+(t)\eta _+(t')
\right]. 
\end{multline}%
As a result we have 
\eqM{ 
T_+ \exp \left[
-i \int dt\, \eta _+(t) \hat q(t)
\right] \\ = \exp \left[ - 
\frac{i\hbar }{2} \int dt dt' D_F(t-t') 
\eta _+(t)\eta _+(t')\right]\, \bm{:}\exp \left[
-i \int dt\,\eta _+(t) \hat q(t)   
\right]  
 \bm{:} 
\, ,  
}{
\eqlabel{WickClosed} 
}%
which is indeed relation (\ref{eq:Wick}) written for the case of 
$T_+$-ordering, i.e., with $\eta _- = 0 $. 

Generalisation of this argument to the double time ordering 
reduces to applying Wick's theorem to the C-contour (Fig.\ \ref{fig:C}) in place of the time axis, cf., e.g., Ref.\ \cite{Keldysh}. Hori's form of Wick's theorem for 
the double time ordering reads (cf.\ \cite{BWO}) 
\eqM{ 
T_- \hat q(t_1)\cdots \hat q(t_m) \, 
T_+ \hat q(t'_1)\cdots \hat q(t'_n) \\ = \bm{:} 
\left.\big[\exp({\Delta_C})\, 
q_-(t_1)\cdots q_-(t_m)
q_+(t'_1)\cdots q_+(t'_n)
\big]\right|_{q_{\pm}\to \hat q} \bm{:} 
\, ,    
}{
\eqlabel{HoriC} 
}%
where 
\eqM{ 
\Delta_C = i\hbar\int dt dt' \bigg[
\frac{1 }{2} D_F(t-t') 
\frac{\delta^2}{\delta q_+(t)\delta q_+(t')}
\\ - \frac{1 }{2} D^*_F(t-t') 
\frac{\delta^2}{\delta q_-(t)\delta q_-(t')} 
+ D(t-t')\frac{\delta^2}{\delta q_-(t)\delta q_+(t')}
\bigg]   
. 
}{
}%
This can verified in detailed analogy to how \Eq{Hori} was proven, cf.\ \Eqs{OneDelta} and (\ref{eq:AllDeltas}). 
The rest of the calculation leading from (\ref{eq:WickHori}) to (\ref{eq:WickClosed}) generalises trivially resulting in \Eq{Wick}. 
\section{The symmetric mapping}\flashlabel{app:Symm} 
Here we give a proof of equation (\ref{eq:WResp}).  
Regarded as functional variables, currents $j$ and $\jW$ are coupled by the substitution 
\begin{gather} 
\begin{aligned}
j(t) = \jW(t) + \frac{i\hbar }{2}\left[
\eta ^{(+)}(t) - \eta ^{(-)}(t)
\right]  . 
\end{aligned}%
\end{gather}%
By making use of \Eqs{FOscFin0} and (\ref{eq:PhiRJ}) we then find 
\begin{gather} 
\begin{aligned}
\Phi \left(
\frac{ j_-}{\hbar } , \frac{ j_+}{\hbar }
\right) = \Phi _{\text{cl}} (\eta ,\jW) \FW(\eta ),  
\end{aligned}%
\eqlabel{FAltW} 
\end{gather}%
where 
\begin{multline} %
\preprintmargin
\FW(\eta ) = \nav{\exp\int dt\, \eta (t) \hat q (t)} 
\,\\ \times 
\exp\bigg\{\frac{i\hbar }{2}
\int dt dt' 
\eta (t) 
D\R(t-t') \Big[
\eta ^{(+)}(t') - \eta ^{(-)}(t')
\Big] 
\bigg\} . 
\eqlabel{FWDef} %
\preprintmargin
\end{multline}%
It is a well-known fact (see, e.g., \cite{AgWo}) that characterisrtic functionals corresponding to different operator orderings differ in Gaussian factors. This is exactly the case for $\FW(\eta )$ and $\Phi _{\text{in}} (\eta )$, so that $\FW(\eta )$ must be a characteristic functional corresponding to some ordering other than the normal ordering. Below we prove that $\FW(\eta )$ is a symmetrically ordered characteristic functional of the initial state, 
\begin{gather} 
\begin{aligned}
\FW(\eta ) = \qav{W \exp\int dt\, \eta (t) \hat q (t)}. 
\end{aligned}%
\eqlabel{FWW} 
\end{gather}%
Equation (\ref{eq:WResp}) then follows by combining \Eqs{FAltW} and (\ref{eq:FWW}) and recalling \Eq{qjbyq}. 

We start the proof of (\ref{eq:FWW}) from simplifying the Gaussian factor in \Eq{FWDef}. Equation (\ref{eq:FPNConv}) allows us to shift the frequency-positive and negative parts on $D\R$, 
\begin{multline} %
\preprintmargin
\int dt dt' \eta (t) D\R(t-t')\left[
\eta ^{(+)}(t') - \eta ^{(-)}(t')
\right]  \\ = \int dt dt' \eta (t) \left[
D^{(+)}\R(t-t') - D^{(-)}\R(t'-t)
\right]\eta (t') . 
\preprintmargin
\end{multline}%
We have also changed $D^{(-)}\R(t-t')\to D^{(-)}\R(t'-t)$; it does not affect the result because of the symmetrisation imposed by the quadratic form. 
Using \Eq{OscDbyDR} we obtain 
\eqA{ 
\FW(\eta ) = \nav{\exp\int dt\, \eta (t) \hat q (t)} 
\exp\left[\frac{i\hbar }{2}
\int dt dt' \eta (t) D(t-t') \eta (t')
\right]  . 
}{
\eqlabel{FWND} %
}%
To prove \Eq{FWW} it suffices to show that quadratic terms in (\ref{eq:FWW}) and (\ref{eq:FWND}) coincide: 
\begin{multline} %
\preprintmargin
\frac{1}{2} \int dt dt' \eta (t) \eta (t') 
\Sbracket{\big}{
 \nav{\hat q(t)\hat q(t')} + i \hbar D(t-t')
} 
\\ = \frac{1}{2} \int dt dt' \eta (t) \eta (t') 
 \qav{W\hat q(t)\hat q(t')}. 
\preprintmargin
\end{multline}%
A nearly identical relation follows directly from the definition of $D(t)$ as contraction by \Eq{DDef}:  
\begin{multline} %
\preprintmargin
\frac{1}{2} \int dt dt' \eta (t) \eta (t') \Sbracket{\big}{
 \nav{\hat q(t)\hat q(t')} + i \hbar D(t-t')
}  \\ = \frac{1}{2} \int dt dt' \eta (t) \eta (t') 
 \qav{\hat q(t)\hat q(t')}. 
\preprintmargin
\end{multline}%
In view of the symmetrisation of kernels imposed by the quadratic form we are left to prove that 
\begin{gather} 
\begin{aligned}
\frac{1}{2}\big[
 \hat q(t)\hat q(t') + \hat q(t')\hat q(t)
\big]  =  W \hat q(t)\hat q(t'). 
\end{aligned}%
\end{gather}%
This is easily checked by making use of \Eq{OscHeis}: 
\begin{multline} %
\preprintmargin
\frac{1}{2}\big[
 \hat q(t)\hat q(t') + \hat q(t')\hat q(t)
\big] = \frac{q_0^2}{4} \big[
\hat a^2 \textrm{e}^{-i\omega _0(t+t')} + 
\hat a\dgp{2} \textrm{e}^{i\omega _0(t+t')} \\ + 
(\hat a \hat a\dg + \hat a\dg \hat a)\cos\omega _0(t-t')
\big] ,  
\preprintmargin
\end{multline}%
in which formula the symmetric ordering is evident. 

\begin{thebibliography}{} 
\bibitem{Schleich}
Wolfgang P.\ Schleich, {\em Quantum Optics in Phase Space\/} 
(Wiley, Berlin, 2001). 
\bibitem{MandelWolf}
Leonard Mandel and Emil Wolf, {\em Optical coherence and quantum optics\/} 
(Cambridge University Press, 1995).
\bibitem{Bogol}
N.N.\ Bogoliubov and D.V.\ Shirkov, 
{\em Introduction to the theory of quantized fields\/} 
(Wiley, New York, 1980). 
\bibitem{Zinn}
J.\	Zinn-Justin, {\em Quantum field theory and critical phenomena\/} 
(Oxford, Clarendon Press, 2004). 
\bibitem{Keldysh}
O.V.\ Konstantinov and V.I.\ Perel, 
Zh.\ Eksp.\ Theor.\ Phys.\ {\bf 39}, 197 (1960) 
[Sov.\ Phys.\ JETP {\bf 12}, 142 (1961)];  
L.V.\ Keldysh, {\em ibid.\/} {\bf 47}, 1515 (1964) 
[{\bf 20}, 1018 (1964)]. 
\bibitem{Nelson}
E.\ Nelson, Journal of Functional Analysis {\bf 12}, 97 (1973). 
\bibitem{Vogel}
Werner Vogel, Dirk-Gunnar Welsch, and Sascha Wallentowitz, 
{\em Quantum Optics. An Introduction\/}
(Wiley 2006). 
\bibitem{QED}
L.I.\ Plimak, Phys.\ Rev.\ A {\bf 50}, 2120 (1994). 
\bibitem{SchwingerJ}
J.S.\ Schwinger,  
Phys.\ Rev.\ {\bf 152}, 1219 (1966); {\em ibid.\/} {\bf 158}, 1391 (1967). 
\bibitem{Kubo}
R.\ Kubo, {\em Statistical Mechanics\/} (North-Holland, 1965).
\bibitem{GlauberPhDet}
Roy J.\ Glauber, Phys.\ Rev.\ {\bf 130}, 2529 (1963). 
\bibitem{SchwingerC}
J.S.\ Schwinger,  
J.\ Math. Phys.\ {\bf 2}, 407 (1961). 
\bibitem{KellyKleiner}
P.L.\ Kelly and W.H.\ Kleiner, Phys.Rev.\ {\bf 136}, A316 (1964).  
\bibitem{GlauberTN}
R.J.\ Glauber, {\em Quantum Optics and Electronics\/}, 
Les Houches Summer School of Theoretical Physics 
(Gordon and Breach, New York, 1965). 
\bibitem{Hori}
T.\ Hori, 
Prog.\ Theor.\ Phys.\ {\bf 7}, 378 (1952). 
\bibitem{Cooper}
Fred Cooper, {\em e-print\/} arXiv:hep-th/950407v1 (1995). 
\bibitem{AgWo}
{G.S. Agarwal and E. Wolf, \PRD {\bf 2}, 2161, (1970).}
\bibitem{BWO}
L.I.\ Plimak, M.\ Fleischhauer, M.K.\ Olsen, and M.J.\ Collett, 
Phys.\ Rev.\ A {\bf 67}, 013812 (2003). 

\bibitem{IntRange}
Whenever integration limits are omitted, a maximal possible range of integration is implied: 
the whole time axis, 
the whole space, and so on.   


\hide{
\bibitem{Feedback}
L.I.\ Plimak, Quant.\ Semicl.\ Opt.\ {\bf 8}, 323 (1996).  
\bibitem{YamFeedB}
S.\ Machida and Y.\ Yamamoto, Opt.\ Comm.\ {\bf 57}, 290 (1986).
\bibitem{PlimakWalls}
L.I.\ Plimak and D.F.\ Walls,
Phys.\ Rev.\ A {\bf 50}, 2627 (1994). 
\bibitem{Prepr}
L.I.\ Plimak, M.\ Fleischhauer, M.K.\ Olsen, and M.J.\ Collett, 
{\em Quantum-field-theoretical techniques for stochastic 
representation of quantum problems\/}, e-print cond-mat/0102483.
\bibitem{EuLett}
L.I.\ Plimak {\em et al\/}., Europhys.\ Lett.\ {\bf 56}, 372 (2001). 
\bibitem{AskMurray1}
M.K.\ Olsen, L.I.\ Plimak, and M.J.\ Collett, 
Phys.\ Rev.\ A {\bf 64}, 063601 (2001). 
\bibitem{AskMurray2}
M.K.\ Olsen, L.I.\ Plimak, and M.\ Fleischhauer, 
Phys.\ Rev.\ A {\bf 65}, 053806 (2002). 
\bibitem{Cumul}
R.L.\ Stratonovich, {\em Nonlinear Nonequilibrium Thermodynamics I\/} 
(Springer, Berlin, 1992).
\bibitem{Steel}
M.J.\ Steel {\em et al.\/}, 
Phys.\ Rev.\ A {\bf 58}, 4824 (1998). 
}
\end{thebibliography}
\end{document}